\documentclass[a4paper, 10pt]{report}

\usepackage[cp1250]{inputenc}
\usepackage[T1]{fontenc}
\usepackage{amsfonts}
\usepackage{graphicx}
\usepackage{amsmath}
\usepackage{caption}

\usepackage{amssymb}

\usepackage{fancyhdr}
\usepackage{bibtopic}
\usepackage{color}
\usepackage{ulem}
\usepackage[toc,page]{appendix}
\usepackage{setspace}
\usepackage{cite}
\usepackage{xcolor}

\AtBeginDocument{}
\usepackage{etoolbox}
\patchcmd{\thebibliography}{\chapter*}{\section*}{}{}

\makeatletter
\renewcommand{\thesection}{%
  \ifnum\c@chapter<1 \@arabic\c@section
  \else \thechapter.\@arabic\c@section
  \fi
}
\makeatother

\patchcmd{\tableofcontents}{\chapter*}{\section*}{}{}

\numberwithin{equation}{section}

\let\OLDthebibliography\thebibliography
\renewcommand\thebibliography[1]{
  \OLDthebibliography{#1}
  \setlength{\parskip}{0pt}
  \setlength{\itemsep}{3.5pt plus 1ex}
}

\usepackage[linktocpage]{hyperref}
\definecolor{darkred}{rgb}{0.5,0,0}
\definecolor{darkpurple}{rgb}{0.5,0,0.5}
\definecolor{darkblue}{rgb}{0,0,0.5}
\hypersetup{ colorlinks,
linkcolor=darkblue,
filecolor=darkpurple,
urlcolor=darkred,
citecolor=darkpurple }
 
\usepackage{soul}

\begin{document}

\allowdisplaybreaks
\setlength{\abovedisplayskip}{3.5pt}
\setlength{\belowdisplayskip}{3.5pt}
\abovedisplayshortskip
\belowdisplayshortskip

{\setstretch{1.0}

{\LARGE \bf \centerline{
Evolution of perturbations in a universe
}}
{\LARGE \bf \centerline{
with exotic solid-like matter
}}

\vskip 1cm
\begin{center}
{Peter M\'esz\'aros\footnote{e-mail address: peter.meszaros@fmph.uniba.sk}}

\vskip 2mm {\it Department of Theoretical Physics, Comenius
University, Bratislava, Slovakia}

\vskip 2mm \today 
\end{center}

\section*{Abstract}

We study cosmological perturbations in a universe with only one matter component described by a triplet of fields. Configuration of these fields is the same as for body coordinates of a solid, and they enter the matter Lagrangian only through the kinetic term. We restrict ourselves only to cases with constant pressure to energy density ratio $w$. Superhorizon perturbations have no constant modes with scalar vector and tensor perturbations decaying or growing at different rates, and in cases with pressure to energy density ratio $w>(19-8\sqrt{7})/3\dot{=}$ $\dot{=}-0.722$ perturbations propagate with superluminal sound speed. Regarding our universe, these results illustrate possible challenges with comparing the observational data to models similar to solid inflation, if the inflation is followed by a period during which the studied model is a sufficiently good approximation.

\vskip 3mm \hspace{4mm}
\begin{minipage}[t]{0.8\textwidth}
\noindent\rule{12cm}{0.4pt}
\vspace{-9mm}
\tableofcontents
\noindent\rule{12cm}{0.4pt}
\end{minipage}
\vskip 6mm

}

\section{Introduction}\label{sec:1}
During the time period between the end of cosmic inflation and the time when nonlinear effects
started to influence the growth of the structure, when perturbations are small,
the matter content of the universe can be considered a multicomponent fluid with viscosity caused by
interaction between its components, \cite{silk}.
The theoretical approach to cosmological perturbations \cite{bardeen,mukhanov} with such description of the matter
filling the universe leads to successful fitting of the $\Lambda$CDM cosmological model to observational data \cite{planck}.
Although the most prominent current problem with the Hubble tension \cite{riess,valentino} remains still unsolved.
\vskip 2mm
In order to either study nonbaryonic components of the $\Lambda$CDM universe or the inflationary period,
forms of matter other than perfect fluid are taken into account.
The simplest and most studied case is single scalar field \cite{ratra,ellis,maartens,barrow}.
It can describe quintessence models of dark energy \cite{weinberg,ferreira,zlatev} through different approaches,
including nonminimal coupling to gravity \cite{fujii,ford,wetterich}, k-essence \cite{chiba,picon1,picon2} and the Chaplygin gas \cite{kamen}.
Scalar field can also drive the inflation \cite{brout,linde,maldacena}, it can be used in models alternative to inflation \cite{buchbinder,ijjas,andrei}, describe dark matter \cite{sin,magana}, or both dark matter and dark energy at the same time \cite{bento,scherrer,mamon}.
A natural extension of single field models is the multifield approach, which is usually studied in inflationary
context \cite{starobinsky,polarski,nakamura,sasaki,gong,lee,gariga,christ,palia,morishita}.
There is no shortage of models generalizing the multifield approach even more, and among them there are
models inspired by general relativistic elasticity \cite{carter} with a triplet of fields playing the same
role as body coordinates of a solid.
Such concept of matter has been studied in general cosmological context \cite{bucher,dubovsky,skovran,balek},
but mostly as the inflationary model \cite{gruzinov,endlich,akhshik,bartolo,bordin} with broad further
development \cite{ricciardone,ja,celoria1,cabass1,celoria2,cabass2,comelli,aoki}.
\vskip 2mm
In this paper, we focus on a model which is a special case of solid models,
but at the same time, it generalizes the single field k-essence to a triplet of fields $\varphi^A$,
so that the matter Lagrangian is of the form
$\mathcal{L}_{\textrm{m}} = f \left( \Sigma_{A=1}^{3} g^{\mu\nu}\varphi^{A}_{\phantom{A},\mu}\varphi^{A}_{\phantom{A},\nu} \right)$.
For simplicity, we assume that this triplet of fields is the only matter component
of the universe, and we adopt the standard perturbation theory \cite{bardeen} with the flat Friedmann--Lema\^itre--Robertson--Walker
(FLRW) background.
When analyzing this model, we have to pay close attention to
two important features of other cosmological models with solid studied so far.
There are cases with superluminal propagation of perturbations \cite{dubovsky},
and superhorizon perturbations are not conserved \cite{endlich}.
\vskip 2mm
The model studied in this paper differs from other works with similar forms of the matter Lagrangian. Unlike in works focused on cosmic inflation \cite{gruzinov,endlich,akhshik,bartolo,bordin,ricciardone,ja,celoria1,cabass1,celoria2,cabass2,comelli,aoki}, we allow the pressure to energy density ration to be significantly different from $w=-1$. In other works dealing with the solid matter in the more general cosmological context \cite{bucher,dubovsky,skovran,balek} there is matter Lagrangian in the form of the Taylor expansion in the terms of traces of the body metric $B^{AB}=g^{\mu\nu}\varphi^{A}_{\phantom{A},\mu}\varphi^{B}_{\phantom{B},\nu}$, and coefficients of this expansion are associated with quantities like Lam\'e coefficients. In the main part of this paper, we will study models with matter Lagrangian proportional to some, in general noninteger, power of $\textrm{Tr}(B)=\Sigma_{A=1}^{3} g^{\mu\nu}\varphi^{A}_{\phantom{A},\mu}\varphi^{A}_{\phantom{A},\nu}$, and therefore, the standard approach with Taylor expansion would not be applicable.
\vskip 2mm
In section \ref{sec:2} we explain the model under consideration
in more detail, and in section \ref{sec:3} we impose a further restriction - constant pressure to energy
density ratio.
The bulk of the paper, sections \ref{sec:4}-\ref{sec:6}, is dedicated to the
first order perturbations, and our results are summarized in the last section \ref{sec:7}.
We use units in which the light speed is $c=1$, and the signature $(-,+,+,+)$ for the spacetime metric.

\section{Matter Lagrangian}\label{sec:2}
The general form of the kinetic term in a field theory with multiple fields labeled by capital
Latin indices is
\begin{eqnarray}\label{eq:kin}
\mathcal{K} \equiv -\dfrac{1}{2}\mathcal{X} =-\dfrac{1}{2}K_{AB}g^{\mu\nu}\varphi^{A}_{\phantom{A},\mu}\varphi^{B}_{\phantom{B},\nu},
\end{eqnarray}
where components of the matrix $K_{AB}$ can be usually chosen as $K_{AB}=\delta_{AB}$
by redefinition of fields. Quantity $\mathcal{X}$ is here defined for the convenience,
since it will be useful throughout the rest of the paper.
In a homogeneous and isotropic universe the most natural choice
for the configuration of fields is to assume that they depend only on time, $\varphi^{A}=\varphi^{A}(\tau)$,
however, in the case with a triplet of fields, one can also set
\begin{eqnarray}\label{eq:fi0}
\varphi^{A}=\alpha\delta^{A}_{i}x^{i},
\end{eqnarray}
where $\alpha$ is a constant.
This is also an isotropic and homogeneous configuration as long as the matter Lagrangian does not
depend on the fields directly, for example through some potential $V(\varphi^1,\varphi^2,...)$.
In other words, the matter Lagrangian may depend only on the kinetic term,
\begin{eqnarray}\label{eq:lag}
\mathcal{L}_{\textrm{m}}=-f(\mathcal{X}), \quad \textrm{where} \quad \mathcal{X}=g^{\mu\nu}\varphi^{i}_{\phantom{i},\mu}\varphi^{i}_{\phantom{i},\nu},
\end{eqnarray}
where for simplicity we dropped capital Latin indices and replaced them with indices denoting spatial coordinates,
and repeating two such indices indicates summation even when both of them are upper. With this convention
we can simply write $\varphi^i=\alpha x^i$ for the background configuration. With the minus sign in (\ref{eq:lag}),
the function $f$ directly represents the energy density.
\vskip 2mm
The most natural choice is $\mathcal{L}_{\textrm{m}}=\mathcal{K}=-(1/2)\mathcal{X}$, i.e. matter Lagrangian of a massless free triplet of fields, however, in our case, the background configuration of fields is given by (\ref{eq:fi0}). The pressure to energy density ratio with this choice is $w=-1/3$, but in order to keep other values of $w$ under consideration, we will keep the general form (\ref{eq:lag}). We will assume an additional restriction on it in the next section. We return to the special case with matter Lagrangian proportional to $\mathcal{X}$ in section \ref{sec:6}, because, as we will see later, it requires separate treatment.
\vskip 2mm
The approach described above is used for general relativistic solid matter, where three fields
$\varphi^A$ are called body coordinates, and body metric $B^{AB}$ is defined through push-forward
of the spacetime metric with respect to map from the spacetime to the body space,
$B^{AB}=$ $=g^{\mu\nu}\varphi^{A}_{\phantom{A},\mu}\varphi^{B}_{\phantom{B},\nu}$,
and it is used for the kinematic description of the solid. The kinetic term (\ref{eq:kin}) up to the
factor $-1/2$ then equals trace of the body metric, $\mathcal{K}=-(1/2)\textrm{Tr}(B)$, or $\mathcal{X}=\textrm{Tr}(B)$.
The matter Lagrangian (\ref{eq:lag}) represents only a special case of solid.
The matter Lagrangian of a solid with homogeneous and isotropic properties is invariant with respect
to global rotational and translational internal symmetries,
\begin{eqnarray}
\varphi^{A}\to R^{A}_{\phantom{A}B}\varphi^{B}+T^{A}, \quad R^{A}_{\phantom{A}B}\in SO(3), \quad T^{A}\in \mathbb{R},
\end{eqnarray}
and it can be a function of not only the trace of the body metric but also a function of traces of
its powers, $\mathcal{L}_{\textrm{m}}=F\left(\textrm{Tr}(B),\textrm{Tr}(B^2),\textrm{Tr}(B^3)\right)$.
In this paper, we restrict ourselves to the special form of the matter Lagrangian (\ref{eq:lag}),
because of the closest resemblance to the usual multifield approach with the kinetic term (\ref{eq:kin}).

\section{Pressure to energy density ratio}\label{sec:3}
In order to study cases with the simplest cosmological expansion, we impose an additional condition - constant
pressure to energy density ratio. We assume that it is constant up to at least the first perturbative order,
which will restrict the form of the function $f$ describing the matter Lagrangian (\ref{eq:lag}).
\vskip 2mm
We will employ the standard perturbation theory with the flat FLRW metric and
fields configuration $\varphi^i=\alpha x^i$ as the background. For the parameterization of the spacetime metric and
matter fields up to the first order we use
\begin{eqnarray}\label{eq:mem}
&& ds^2=a(\tau)^2\left\{-(1+2\phi)d\tau^2 + 2 S_i d\tau dx^i + \left[(1-2\psi)\delta_{ij}+\gamma_{ij}\right] dx^i dx^j \right\}, \nonumber\\
&& \varphi^i = \alpha \left( x^i + \zeta_{,i} + \xi_{\perp i} \right).
\end{eqnarray}
For the scalar part of metric perturbations we have chosen the longitudinal gauge, and it is parameterized by functions $\phi$ and $\psi$.
Convenience of this choice is that the gauge invariant perturbations $\widetilde{\phi}$ and $\widetilde{\psi}$ can be expressed simply as $\widetilde{\phi}=\phi$ and $\widetilde{\psi}=\psi$,
and the same is true also for the gauge invariant energy density perturbation.
For the vector part of metric perturbations, we use the gauge with no vector contribution to $g_{ij}$, so that it is parameterized by $S_i$ for which $S_{i,i}=0$,
and for the tensor part we have $\gamma_{ij}$ satisfying conditions $\gamma_{ii}=0$ and $\gamma_{ij,j}=0$.
Both vector and tensor metric perturbations defined in this way are gauge invariant as well, $\widetilde{S}_i=S_i$, $\widetilde{\gamma}_{ij}=\gamma_{ij}$.
Similarly, perturbations of the matter fields are decomposed into the scalar part $\zeta$,
and the vector part $\xi_{\perp i}$ satisfying $\xi_{\perp i,i}=0$.
\vskip 2mm
For the matter Lagrangian (\ref{eq:lag}), the stress-energy tensor can be obtained through the canonical formula
\begin{eqnarray}
T_{\mu\nu}=-2\dfrac{\partial\mathcal{L}_{\textrm{m}}}{\partial g^{\mu\nu}}+\mathcal{L}_{\textrm{m}}g_{\mu\nu}
=2f^{\prime}\varphi^{i}_{\phantom{i},\mu}\varphi^{i}_{\phantom{i},\nu}-fg_{\mu\nu},
\end{eqnarray}
and the direct calculation up to the first order of the perturbation theory
with the parameterization given by (\ref{eq:mem}) yields
\begin{eqnarray}\label{eq:stress}
&& T_{00} = a^2 f \left[ 1 + 2\phi + \dfrac{1}{3}\dfrac{f^{\prime}\mathcal{X}}{f} \left( 6 \psi + 2 \triangle \zeta \right) \right], \nonumber\\
&& T_{0i} = a^2 f \left[ - S_i + \dfrac{2}{3}\dfrac{f^{\prime}\mathcal{X}}{f} \left( \zeta_{,i} + \xi_{\perp i} \right)^{\prime} \right], \nonumber\\
&& T_{ij} = a^2 f \bigg\{ \left[ -1 + \dfrac{2}{3}\dfrac{f^{\prime}\mathcal{X}}{f} + 2 \psi + \dfrac{1}{3}\dfrac{f^{\prime}\mathcal{X}}{f} \left( - 1 + \dfrac{2}{3} \dfrac{f^{\prime\prime}\mathcal{X}}{f^{\prime}} \right) \left( 6 \psi + 2 \triangle \zeta \right) \right] \delta_{ij} + \nonumber\\
&& \phantom{ T_{ij} = a^2 f\bigg\{ } + \dfrac{2}{3}\dfrac{f^{\prime}\mathcal{X}}{f} \left( \xi_{\perp i,j} + \xi_{\perp j,i} + 2 \zeta_{,ij} \right) - \gamma_{ij} \bigg\}.
\end{eqnarray}
Here and from now on the prime plays two different roles. It denotes partial derivative with respect to $\mathcal{X}$ when it is at $f$,
so that $f^{\prime}=\partial f/\partial\mathcal{X}$, and derivative with respect to conformal time when it is at any perturbation,
for example $\phi^{\prime}=d\phi/d\tau$.
Note also that $f$, $f^{\prime}$, $f^{\prime\prime}$ and $\mathcal{X}$ in relations (\ref{eq:stress}) are evaluated at
the background configuration $\mathcal{X}=3(\alpha/a)^2$.
\vskip 2mm
As we can see in the last line of (\ref{eq:stress}), there are nonzero shear stress components of the stress-energy tensor.
This leads to behavior of perturbations which is different from more standard cases of matter filling the universe,
for example, perfect fluid or single scalar field, which will be the main topic of this paper. But for now, we focus only
on the energy density and pressure parts. In the case with a perfect fluid, one can extract them from the stress-energy tensor
$T_{\mu\nu}=(\rho+p)u_{\mu}u_{\nu}+pg_{\mu\nu}$ with $u^{\mu}$ denoting the 4-velocity of volume elements, through relations
\begin{eqnarray}
\rho = - T_{0}^{\phantom{0}0}, \quad p = \dfrac{1}{3} T_{i}^{\phantom{i}i},
\end{eqnarray}
which are valid even up to the first perturbative order. By adopting these relations as definitions of energy density and pressure for matter studied in this paper we obtain
\begin{eqnarray}\label{eq:pom1}
&& \rho = f \left[ 1 + \dfrac{1}{3}\dfrac{f^{\prime}\mathcal{X}}{f} \left( 6 \psi + 2 \triangle \zeta \right) \right], \nonumber\\
&& p = f \left[ -1 + \dfrac{2}{3}\dfrac{f^{\prime}\mathcal{X}}{f} + \dfrac{1}{3}\dfrac{f^{\prime}\mathcal{X}}{f} \left( - \dfrac{1}{3} + \dfrac{2}{3} \dfrac{f^{\prime\prime}\mathcal{X}}{f^{\prime}} \right) \left( 6 \psi + 2 \triangle \zeta \right) \right].
\end{eqnarray}
Thanks to the same combination of perturbations appearing in both pressure and energy density,
$6\psi+2\triangle\zeta$, it is possible to demand the pressure to energy density ratio to be
constant even with perturbations.
\vskip 2mm
The background pressure to energy density ratio derived from (\ref{eq:pom1}) is
\begin{eqnarray}\label{eq:wratio}
\dfrac{\overline{p}}{\overline{\rho}} = -1 + \dfrac{2}{3}\dfrac{f^{\prime}\mathcal{X}}{f},
\end{eqnarray}
and it can be constant only if the function $f$ satisfies the equation $f^{\prime}\mathcal{X}=\mathcal{B}f$
with $\mathcal{B}$ being a constant. This reduces its form to $f\left(\mathcal{X}\right)=\mathcal{C} \mathcal{X}^{\mathcal{B}}$,
where $\mathcal{C}$ denotes another constant, and for the background pressure to energy density ratio we have
$\overline{w}=-1+(2/3)\mathcal{B}$.
From relations (\ref{eq:pom1}) we can express also the ratio of the first order pressure perturbation
to the first order energy density perturbation,
\begin{eqnarray}
\dfrac{\delta p}{\delta \rho} = - \dfrac{1}{3} + \dfrac{2}{3} \dfrac{f^{\prime\prime}\mathcal{X}}{f^{\prime}}.
\end{eqnarray}
By using the form of the function $f$ given by the requirement of constant background pressure to energy
density ratio $\overline{w}$, which is $f\left(\mathcal{X}\right)=\mathcal{C} \mathcal{X}^{\mathcal{B}}$,
we find $\delta p / \delta \rho = -1/3 + (2/3)\left(\mathcal{B}-1\right)=\overline{w}$. This means that
constant $\overline{w}$ implies constant pressure to energy ratio also with first order perturbations
taken into account.
\vskip 2mm
We can also see that the case with $\mathcal{B}=0$ corresponding to constant matter Lagrangian is exceptional.
In this case, (\ref{eq:pom1}) implies $w=-1$ with disappearing energy density and pressure perturbations.
Therefore, the model studied in this paper is not suitable for describing dark energy because it can be replaced with the ordinary model with the cosmological constant.
However, there are other multifield models generalizing the approach from this paper that allow for dark energy with perturbations \cite{celoria3,comelli2}.
\vskip 2mm
In conclusion, pressure to energy density ratio up to the first perturbative order
is allowed to be constant only if the matter Lagrangian is of the form
\begin{eqnarray}\label{eq:XB}
\mathcal{L}_{\textrm{m}} = - \mathcal{C} \mathcal{X}^{\mathcal{B}},
\end{eqnarray}
with arbitrary constants $\mathcal{C}$ and $\mathcal{B}$,
and the corresponding pressure to energy ratio is
\begin{eqnarray}
w = - 1 + \dfrac{2}{3} \mathcal{B}.
\end{eqnarray}
From now on we restrict ourselves to cases with mater Lagrangian of this form.

\section{Radiation-like case}\label{sec:4}
\begin{figure}[!htb]
\sbox0{
\begin{tabular}{rl}
\includegraphics[scale=0.7]{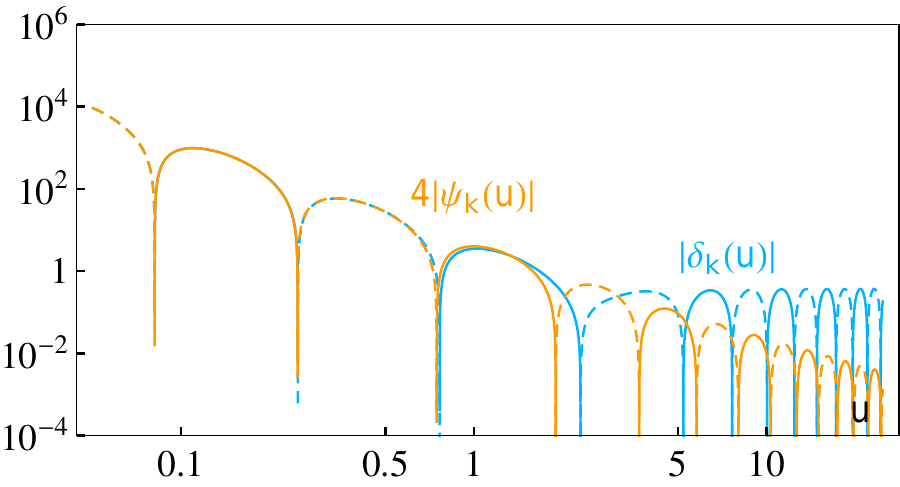} &
\includegraphics[scale=0.7]{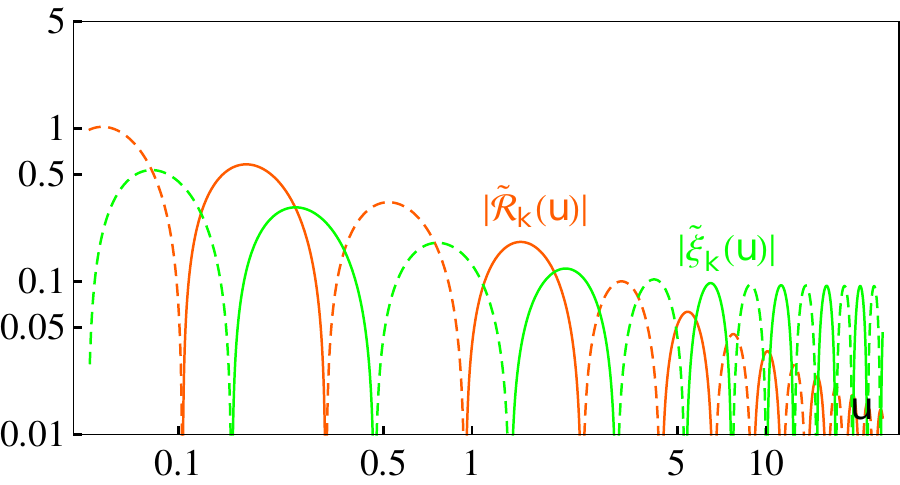} \\
\hline
\begin{tabular}{r}
{\scriptsize The case with perfect fluid radiation:} \\
\includegraphics[scale=0.3]{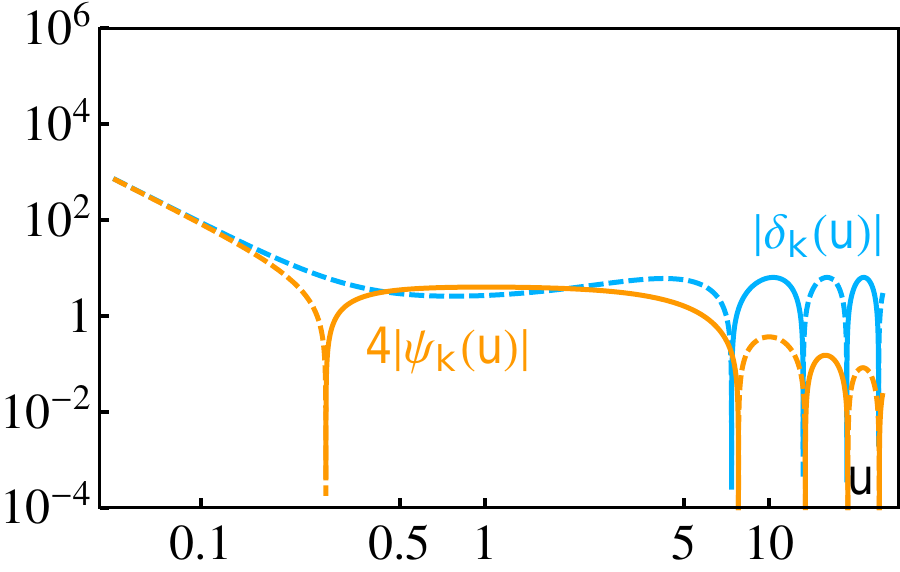}
\includegraphics[scale=0.3]{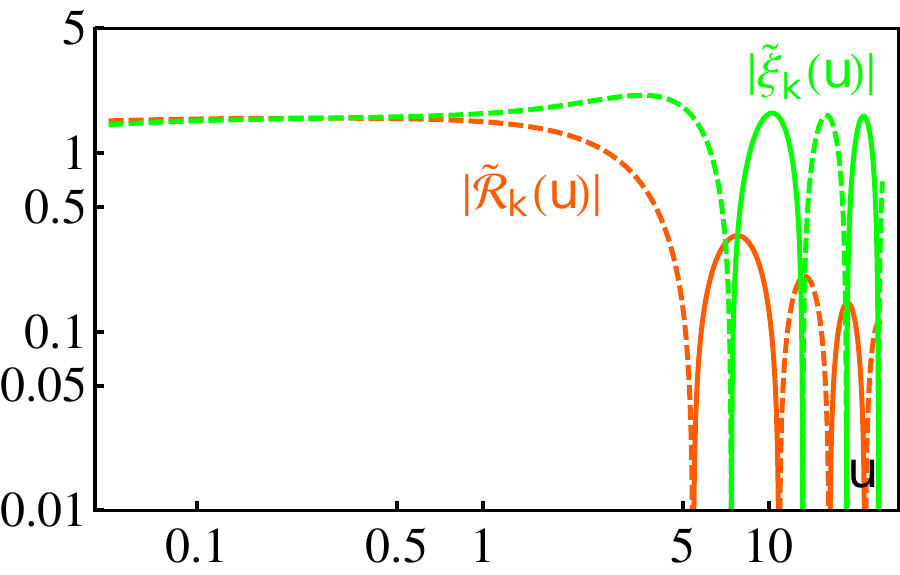}
\end{tabular}
&
{\tiny $\begin{array}{l}
\dfrac{d^2\psi_{k}}{du^2}+\dfrac{4}{u}\dfrac{d\psi_{k}}{du}+\dfrac{1}{3}\psi_{k}=0 \\
\delta_{k} = -2\left(1+\dfrac{1}{3}u^2\right)\psi_k-2u\dfrac{d\psi_{k}}{du} \\
\widetilde{\mathcal{R}}_{k}=-\dfrac{3}{2}\psi_k-\dfrac{1}{2}u\dfrac{d\psi_k}{du} \\
\widetilde{\mathcal{\xi}}_{k}=-\dfrac{3}{2}\left(1+\dfrac{1}{9}u^2\right)\psi_k-\dfrac{1}{2}u\dfrac{d\psi_k}{du}
\end{array} $}
\end{tabular}
}
\centering
\begin{minipage}{\wd0}
\usebox0
\linespread{1}
\setlength{\abovecaptionskip}{-2pt plus 0pt minus 0pt}
\caption{\label{fig:02}{\footnotesize
Numerical solutions of the equation (\ref{eq:13psi}) with conditions $\psi_{k}(1)=1$
and $\psi_{k}^{\prime}(1)=0$ together with quantities expressed through relations (\ref{eq:rest1})-(\ref{eq:rest3}).
For easy comparison, in the bottom part under the line, we plot also the solution with the same conditions,
$\psi_{k}(1)=1$ and $\psi_{k}^{\prime}(1)=0$, in the case with radiation described as perfect fluid,
and we also write down corresponding equations.
Orange lines correspond to $4\psi_{k}$, blue lines to $\delta_{k}$,
red lines to $\widetilde{\mathcal{R}}_{k}$, and green lines to $\widetilde{\xi}_{k}$.
Horizontal axes for all plots correspond to quantity $u$.
We use a logarithmic plot and we depict the absolute values of modes with the conditions mentioned
above. Positive and negative values of these modes are indicated by solid and dashed lines respectively.
}}
\end{minipage}
\end{figure}

\begin{figure}[!htb]
\sbox0{
\includegraphics[scale=0.7]{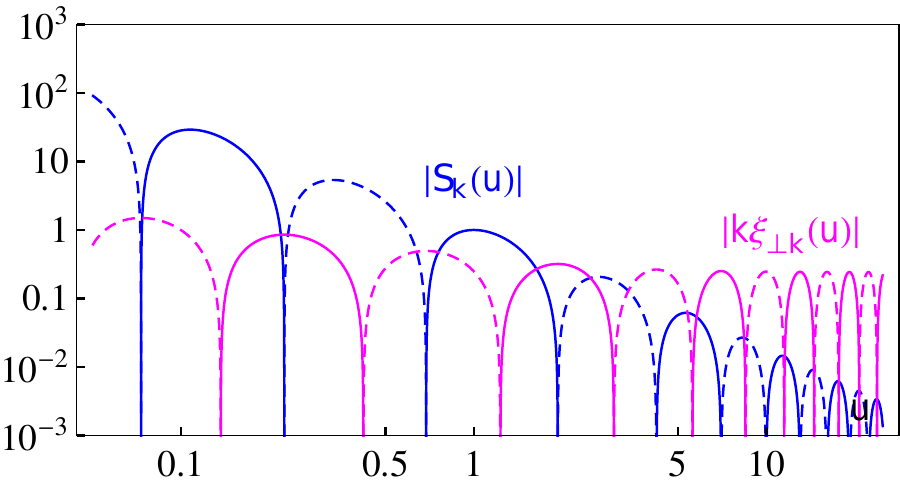}
\hspace{2mm}
\includegraphics[scale=0.7]{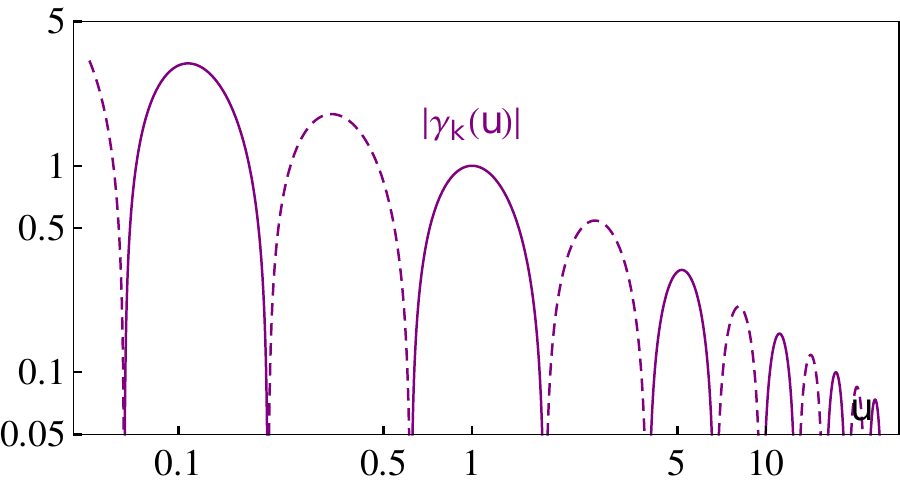}
}
\centering
\begin{minipage}{\wd0}
\usebox0
\linespread{1}
\setlength{\abovecaptionskip}{-10pt plus 0pt minus 0pt}
\caption{\label{fig:03}{\footnotesize
Modes of vector and tensor perturbations given by solutions (\ref{eq:13ss}) and (\ref{eq:13tt}) with condition $S_{k}(1)=1$, $S^{\prime}_{k}(1)=0$, $\gamma_{k}(1)=1$ and $\gamma^{\prime}_{k}(1)=0$,
and with the same conventions as in Fig. \ref{fig:02}. The absolute value of $S_{k}(u)$ is depicted by blue line,
$\xi_{\perp k}(u)$ by magenta, and $\gamma_{k}(u)$ by purple line.
}}
\end{minipage}
\end{figure}

In this section, we restrict ourselves to the special case with pressure to energy density ratio $w=1/3$,
the same as for radiation. This corresponds to $\mathcal{B}=2$, so that the matter Lagrangian is
$\mathcal{L}_{\textrm{m}}=-\mathcal{C}\mathcal{X}^2$.
\vskip 2mm
Components of the background stress-energy tensor $\overline{T}_{\mu\nu}$ corresponding the relaxed
configuration $\overline{\varphi}^{i}=\alpha x^i$ with $\overline{\mathcal{X}}=3\alpha^2 a^{-2}$, are
\begin{eqnarray}\label{eq:stress130}
\overline{T}_{00} = \dfrac{9\mathcal{C}\alpha^4}{a^2}, \quad
\overline{T}_{0i} = 0, \quad \overline{T}_{ij} = \dfrac{3\mathcal{C}\alpha^4}{a^2} \delta_{ij}, 
\end{eqnarray}
and the Einstein field equations read
\begin{eqnarray}\label{eq:stress130}
3 \mathcal{H}^2 = 8\pi\kappa \dfrac{9\mathcal{C}\alpha^4}{a^2}, \quad -\mathcal{H}^2-2\mathcal{H}^{\prime} = 8\pi\kappa \dfrac{3\mathcal{C}\alpha^4}{a^2},
\end{eqnarray}
where $\mathcal{H}=a^{\prime}/a$ with prime denoting differentiation with respect to the conformal time $\tau$.
The solution of these equations is $a(\tau)=\sqrt{24\pi\kappa\mathcal{C}\alpha^4}\tau$, where the time $\tau=0$ corresponds to the big bang limit $a\to 0$,
so that $\mathcal{H}=1/\tau$, $\mathcal{H}^{\prime}=-1/\tau^2$.
Due to the Bianchi identity, conservation laws $\overline{T}^{\mu\nu}_{\phantom{\mu\nu};\nu}=0$ yield no equation
independent from the Einstein equations.
Here $\overline{T}^{i\nu}_{\phantom{i\nu};\nu}=0$ is satisfied automatically, and $\overline{T}^{0\nu}_{\phantom{0\nu};\nu}=0$
is equivalent to the equation $\overline{\rho}^{\prime}+4\mathcal{H}\overline{\rho}=0$, from which follows
$\overline{\rho}\propto a^{-4}$, which is satisfied as well, since $\overline{\rho}=-\mathcal{C}\overline{\mathcal{X}}^2$,
and $\overline{\mathcal{X}}=3\alpha^2 a^{-2}$.
\vskip 2mm
For perturbations, we use the parameterization (\ref{eq:mem}).
The perturbed part of the stress-energy tensor is then given by
\begin{eqnarray}\label{eq:stress13}
&& \delta T_{00} = \dfrac{9\mathcal{C}\alpha^4}{a^2} \left[ 2\left(\phi + 2\psi\right) + \dfrac{4}{3} \triangle\zeta \right], \quad
\delta T_{0i} = \dfrac{9\mathcal{C}\alpha^4}{a^2} \left[ - S_i + \dfrac{4}{3} \left( \zeta_{,i} + \xi_{\perp i} \right)^{\prime} \right], \nonumber\\
&& \delta T_{ij} = \dfrac{9\mathcal{C}\alpha^4}{a^2} \left[ \left( \dfrac{2}{3}\psi - \dfrac{4}{9}\triangle\zeta \right) \delta_{ij} + \dfrac{4}{3} \left( 2 \zeta_{,ij} + \xi_{\perp i,j} + \xi_{\perp j,i} \right) - \gamma_{ij} \right].
\end{eqnarray}
Now we can write down Einstein equations for perturbations, and then solve them.
The evolution of the first order perturbations is governed by linear equations. Thus we can treat scalar, vector, and tensor perturbations separately.
\vskip 2mm
For the scalar sector, we have
\begin{eqnarray}
\label{eq:131}
\delta G^{(\textrm{S})}_{00} = 8\pi\kappa \delta T^{(\textrm{S})}_{00} & \implies &
-3\tau\psi^{\prime} + \tau^2\triangle\psi = 3\phi + 6\psi + 2\triangle\zeta, \\
\label{eq:132}
\delta G^{(\textrm{S})}_{0i} = 8\pi\kappa \delta T^{(\textrm{S})}_{0i} & \implies &
\tau^2\psi^{\prime} + \tau\phi = 2\zeta^{\prime}, \\
\label{eq:133}
\delta G^{(\textrm{S})}_{ij} \stackrel{i=j}{=} 8\pi\kappa \delta T^{(\textrm{S})}_{ij} & \implies &
3\tau^2\psi^{\prime\prime} + 3\tau(\phi+2\psi)^{\prime} - 3(\phi+2\psi) + \nonumber\\
 & & + \tau^2\triangle(\phi-\psi) = 2\triangle\zeta, \\
\label{eq:134}
\delta G^{(\textrm{S})}_{ij} \stackrel{i \neq j}{=} 8\pi\kappa \delta T^{(\textrm{S})}_{ij} & \implies &
\tau^2(\psi-\phi) = 8\zeta.
\end{eqnarray}
These equations have been simplified with the use of the background solution.
From (\ref{eq:134}) we can express $\phi$ through $\psi$ and $\zeta$,
and by inserting its expression into the other three equations we obtain
\begin{eqnarray}
\label{eq:135}
(\ref{eq:131}) & \implies & 3\tau\psi^{\prime} - \tau^2\triangle\psi + 9\psi = 24\tau^{-2}\zeta - 2\triangle\zeta, \\
\label{eq:136}
(\ref{eq:132}) & \implies & \tau\psi^{\prime} + \psi = 8\tau^{-2}\zeta + 2\tau^{-1}\zeta^{\prime}, \\
\label{eq:137}
(\ref{eq:133}) & \implies & 3\tau^2\psi^{\prime\prime} + 9\tau\psi^{\prime} - 9\psi = -72\tau^{-2}\zeta + 24\tau^{-1}\zeta^{\prime} + 10\triangle\zeta,
\end{eqnarray}
and after extracting $\zeta^{\prime}$ from (\ref{eq:136}) we can rewrite other equations as
\begin{eqnarray}
\label{eq:138}
(\ref{eq:135}) & \implies & 3\tau\psi^{\prime} - \tau^2\triangle\psi + 9\psi = 2\tau^{-2}(12 - \triangle)\zeta, \\
\label{eq:139}
(\ref{eq:137}) & \implies & -\tau^2\psi^{\prime\prime} + \tau\psi^{\prime} + 7\psi = 2\tau^{-2}(28-5\triangle/3)\zeta.
\end{eqnarray}
Finally, these two equations can be combined to obtain
\begin{eqnarray}\label{eq:13psi}
\dfrac{d^2\psi_k}{du^2} + 4\dfrac{18+u^2}{12+u^2}\dfrac{1}{u}\dfrac{d\psi_k}{du} + \dfrac{1}{3}\dfrac{504+108u^2+5u^4}{12+u^2}\dfrac{1}{u^2}\psi_k = 0,
\end{eqnarray}
where $u=k\tau$ with $k$ denoting the comoving wavenumber, so that the equation above applies to the Fourier mode of the perturbation
$\psi$ with comoving wavevectors $k_i$ such that $k_ik_i=$ $=k^2$.
The convenience of using dimensionless quantity $u$ instead of conformal time $\tau$ is that equations for perturbations rewritten through $u$ have the same form for all modes with various values of the wavenumber. The dimensionless time $u$ can be interpreted as the conformal time measured in multiples of conformal time at which a given mode with wavenumber $k$ crosses the Hubble horizon.
\vskip 2mm
The solution of the equation (\ref{eq:13psi}) can be used to express modes of two remaining perturbations, $\phi$ and $\zeta$ as well.
From (\ref{eq:135}) follows
\begin{eqnarray}\label{eq:13zeta}
k^2\zeta_k = \dfrac{1}{2}\dfrac{u^2}{12+u^2} \left[ (9+u^2)\psi_k + 3u \dfrac{d\psi_k}{du} \right],
\end{eqnarray}
and by using this equation together with (\ref{eq:134}) we find
\begin{eqnarray}\label{eq:13phi}
\phi_k = -\dfrac{3}{12+u^2} \left[ (8+u^2)\psi_k + 4u\dfrac{d\psi_k}{du} \right],
\end{eqnarray}
which concludes the list of all scalar perturbations.
\vskip 2mm
Note that our analysis started with four equation (\ref{eq:131})-(\ref{eq:134}) derived from all potentially
independent components of the Einstein equations, while the number of variables is only three.
However, by extracting $\zeta$ from (\ref{eq:135}), i.e. expressing it through only $\psi$ and $\psi^{\prime}$, inserting it into (\ref{eq:136}),
and replacing $\psi^{\prime\prime}$ with terms given by $\psi$ and $\psi^{\prime}$ with the use of (\ref{eq:13psi}),
we can easily check that (\ref{eq:136}) as equation for $\psi$ and $\psi^{\prime}$ is satisfied automatically.
This means that the equations which we have used are not independent and equations (\ref{eq:13psi})-(\ref{eq:13phi})
contain all information about the dynamics of all independent first order scalar perturbations.
\vskip 2mm
Other useful quantities are gauge invariant fractional perturbation of the energy density
$\delta\equiv{\widetilde{\delta \rho}}/{\overline{\rho}} = {\widetilde{\delta T}_{0}^{\phantom{0}0}}/{\overline{T}_{0}^{\phantom{0}0}}$,
and gauge invariant curvature perturbation which can be parameterized by two quantities
\begin{eqnarray}
\widetilde{\mathcal{R}} = - \widetilde{\psi} + H \delta v, \quad \widetilde{\xi} = - \widetilde{\psi} + \dfrac{\widetilde{\delta\rho}}{3(\overline{\rho}+\overline{p})},
\end{eqnarray}
where $H=\dot{a}/a$ is the standard Hubble parameter, or $H=\mathcal{H}/a$, and $\delta v$ is the potential of matter velocity.
In the model studied here, it can be expressed as $\delta v = - a \zeta^{\prime}$, which can be obtained either from the stress-energy tensor or simply by differentiating the relation $x^{i}=\alpha^{-1}\varphi^{i}-\zeta_{,i}$. The straightforward calculation leads to
\begin{eqnarray}
\label{eq:rest1}
&& \delta_k \equiv \dfrac{\widetilde{\delta \rho}_k}{\overline{\rho}} = 4\psi_k - \dfrac{4}{3}k^2\zeta_k = \dfrac{2}{12+u^2} \left[ \left(24-u^2-\dfrac{1}{3}u^4\right)\psi_k-u^3\dfrac{d\psi_k}{du} \right], \\
\label{eq:rest2}
&& \widetilde{\mathcal{R}}_k = -\psi_k - \mathcal{H}\zeta_k^{\prime} = \dfrac{1}{2}\dfrac{u^2}{12+u^2} \left[ \psi_k - u\dfrac{d\psi_k}{du} \right], \\
\label{eq:rest3}
&& \widetilde{\xi}_k = -\dfrac{1}{3}k^2\zeta_k = -\dfrac{1}{6}\dfrac{u^2}{12+u^2} \left[ \left(9+u^2\right)\psi_k + 3u\dfrac{d\psi_k}{du} \right].
\end{eqnarray}
In the limit of wavenumber of the modes being much larger than the Hubble horizon, $1/k\gg \tau$ or $u\ll 1$, we have
$\delta=4\psi$.
Gauge invariant curvature perturbation is usually  conserved in the superhorizon limit, and therefore, it can be used
for comparing the information encoded by the primordial perturbations, which are stretched to the superhorizon scale during the inflation,
with the observational data. However, as we will see below, this is not the case with our model, which is a problematic issue.
\vskip 2mm
Equation (\ref{eq:13psi}) is too complicated to be solved analytically.
In the superhorizon limit, $u \ll 1$, the equation reduces to
\begin{eqnarray}
\dfrac{d^2\psi_k}{du^2} + \dfrac{6}{u}\dfrac{d\psi_k}{du} + \dfrac{14}{u^2}\psi_k = 0,
\end{eqnarray}
with the general solution of the form
\begin{eqnarray}\label{eq:13psisup}
\psi_k(u) = u^{-5/2}\left[c_1\cos\left( \frac{1}{2}\sqrt{31} \ln u\right)+c_2\sin\left( \frac{1}{2}\sqrt{31} \ln u\right)\right],
\end{eqnarray}
where we use the natural logarithm, and constants $c_1$ and $c_2$ can be fixed by matching initial conditions.
In the subhorizon limit, $u \gg 1$, the equation reduces to
\begin{eqnarray}
\dfrac{d^2\psi_k}{du^2} + \dfrac{4}{u}\dfrac{d\psi_k}{du} + \dfrac{5}{3}\psi_k = 0,
\end{eqnarray}
and its general solution is
\begin{eqnarray}\label{eq:13psisub}
\psi_k(u) & = & \mathcal{U}^{-3/2} \left[ \tilde{c}_3 J_{3/2} \left(\mathcal{U}\right) + \tilde{c}_4 Y_{3/2} \left(\mathcal{U}\right) \right] = \nonumber\\
& = & \mathcal{U}^{-2}\left[ c_3 \left( \cos\mathcal{U} - \dfrac{\sin\mathcal{U}}{\mathcal{U}} \right) + c_4 \left( \sin\mathcal{U} + \dfrac{\cos\mathcal{U}}{\mathcal{U}} \right) \right],
\end{eqnarray}
where $J$ and $Y$ are Bessel functions, $\mathcal{U}=\sqrt{5/3}u=\sqrt{5/3}k\tau$, and
$c_3=-\sqrt{2/\pi}\tilde{c}_3$ and $c_4=-\sqrt{2/\pi}\tilde{c}_4$ are constants
given by initial conditions.
The problematic property of scalar perturbations is the superluminal sound speed,
$c_{\textrm{s}}^{(\textrm{S})2}=5/3$ for $u\to\infty$.
\vskip 2mm
Amplitudes of perturbations are approximatively proportional to some power of $u\propto\tau\propto a$
in both superhorizon and subhorizon limits.
Denote amplitudes of modes of any perturbation $\chi$ as $\mathcal{A}^{(n)}[\chi]$, so that mode of such perturbation corresponding to its $(n)$-th independent solution is the product of its amplitude and function $\mathcal{O}^{(n)}_{k}(u)$
which either oscillates with a constant amplitude or is a constant.
Mode of arbitrary perturbation $\chi$ then can be written as
\begin{eqnarray}\label{eq:defi1}
\chi_{k}(u) = \sum\limits_{n}\mathcal{A}^{(n)}[\chi](a)\mathcal{O}^{(n)}_{k}(u),
\end{eqnarray}
where amplitudes can be expressed in the form of powers of the scale factor as
\begin{eqnarray}\label{eq:defi2}
\mathcal{A}^{(n)}[\chi](a)=\left\{
\begin{array}{ll}
a^{P_0^{(n)}[\chi]} & \textrm{for} \quad u\to 0, \\
a^{P_{\infty}^{(n)}[\chi]} & \textrm{for} \quad u\to \infty.
\end{array}
\right.
\end{eqnarray}
With such conventions, the behavior of superhorizon scalar perturbations is given by
$P^{(n)}_{0}[\psi]=$ $=P^{(n)}_{0}[\delta]=-5/2$, ($\delta\approx4\psi$),
$P^{(n)}_{0}[\widetilde{\mathcal{R}}]=P^{(n)}_{0}[\widetilde{\xi}]=-1/2$,
and in the subhorizon limit, we have
$P^{(n)}_{\infty}[\psi]=-2$, $P^{(n)}_{\infty}[\delta]=0$,
$P^{(n)}_{\infty}[\widetilde{\mathcal{R}}]=-1$ and $P^{(n)}_{\infty}[\widetilde{\xi}]=0$.
As shown in Fig. \ref{fig:02}, the numerical solution is in agreement with this asymptotic behavior.
The second part of the figure under the line is dedicated to the case with perturbations
in a universe filled with radiation described as a perfect fluid.
In such case
$P^{(1)}_{0}[\psi]=P^{(1)}_{0}[\delta]=0$,
$P^{(2)}_{0}[\psi]=P^{(2)}_{0}[\delta]=-3$ and
$P^{(n)}_{0}[\widetilde{\mathcal{R}}]=P^{(n)}_{0}[\widetilde{\xi}]=0$ in the superhorizon limit,
and
$P^{(n)}_{\infty}[\psi]=P^{(n)}_{\infty}[\widetilde{\mathcal{R}}]=-2$, $P^{(n)}_{\infty}[\delta]=P^{(n)}_{\infty}[\widetilde{\xi}]=0$
in the subhorizon limit.
Note also that a significant feature of the solid-like model studied in this paper, as well as
other models with solid matter, is that
$\phi\neq\psi$, whereas in the case with perfect fluid $\phi=\psi$.
This stems from equation (\ref{eq:134}) which yields $\phi=\psi$ in the perfect fluid case,
since $\delta T^{(\textrm{S})}_{ij} \stackrel{i \neq j}{=} 0$ because of vanishing shear stress.
The list of equations for the vector part of perturbations is
\begin{eqnarray}
\label{eq:13s1}
\delta G^{(\textrm{V})}_{0i} = 8\pi\kappa \delta T^{(\textrm{V})}_{0i} & \implies & \left( 2 - \tau^2\triangle \right) S_i = - 6 S_i + 8 \xi_{\perp i}^{\prime}, \\
\label{eq:13s2}
\delta G^{(\textrm{V})}_{ij} = 8\pi\kappa \delta T^{(\textrm{V})}_{ij} & \implies & - 2 \tau S_{(i,j)} - \tau^2 S_{(i,j)}^{\prime} = 8 \xi_{\perp(i,j)}.
\end{eqnarray}
Vector perturbations can be decomposed into modes with two independent polarizations $(+,-)$ as
$S_{ki}(u)=\varepsilon^{+}_{ki}S^{+}_{k}(u)+\varepsilon^{-}_{ki}S^{-}_{k}(u)$, where polarization vectors satisfy $k^{i}\varepsilon^{\pm}_{ki}=0$.
For simplicity, we skip the superscript for the mode functions indicating the polarization,
because both modes $S^{+}_{k}(u)$ and $S^{-}_{k}(u)$ obey the same equation.
From now on, both of these modes will be denoted simply as $S_k(u)$.
Equations (\ref{eq:13s1}) and (\ref{eq:13s2}) can be refined into the more convenient form,
\begin{eqnarray}
\label{eq:13s3}
&& \dfrac{d^2 S_{k}}{du^2} + \dfrac{4}{u}\dfrac{d S_{k}}{du} + \left(1+\dfrac{10}{u^2}\right)S_k = 0, \\
\label{eq:13s4}
&& k\xi_{\perp k} = -\dfrac{1}{4} u \left( S_{k} + \dfrac{1}{2} u \dfrac{S_{k}}{du} \right),
\end{eqnarray}
where $u$ is defined in the same way as before.
The analytic solution of (\ref{eq:13s3}) is given by
\begin{eqnarray}
\label{eq:13ss}
S_{k}(u) = u^{-3/2}\left[c_5 \textrm{Re}\left\{J_{\sqrt{31}i/2}(u)\right\}+ c_6 \textrm{Re}\left\{Y_{\sqrt{31}i/2}(u)\right\} \right].
\end{eqnarray}
Since the order of Bessel functions is imaginary here, this analytic formula
does not provide much insight into the qualitative behavior of modes $S_{k}$.
The approximative behavior is much simpler,
\begin{eqnarray}
\label{eq:13ssa}
S_{k}(u) \approx \left\{ \begin{array}{ll}
u^{-3/2} \left[c_7 \cos\left(\dfrac{1}{2}\sqrt{31}\ln u\right) + c_8 \sin\left(\dfrac{1}{2}\sqrt{31}\ln u\right) \right], & \textrm{for} \quad u\ll 1, \\
u^{-2} \left[c_9 \left(\cos{u}-\dfrac{\sin{u}}{u}\right) + c_{10} \left(\sin{u}+\dfrac{\cos{u}}{u}\right) \right], & \textrm{for} \quad u\gg 1.
\end{array} \right.
\end{eqnarray}
Note that the subhorizon approximation formula, $u\gg 1$, is given by Bessel functions of order $3/2$.
Solution (\ref{eq:13ss}) together with the mode of perturbation $\xi_{\perp}$ given by (\ref{eq:13s4})
for the specific choice of initial conditions is plotted in the left panel of Fig. \ref{fig:03}.
\vskip 2mm
The equation for the tensor part of perturbations can be derived from
\begin{eqnarray}
\label{eq:13t1}
\delta G^{(\textrm{T})}_{ij} = 8\pi\kappa \delta T^{(\textrm{T})}_{ij} & \implies & \left( 2 - \tau^2\triangle \right) \gamma_{ij} + 2\tau\gamma_{ij}^{\prime} + \tau^2\gamma_{ij}^{\prime\prime} = - 6 \gamma_{ij}.
\end{eqnarray}
There are two independent polarizations of tensor modes $(+,\times)$ given by polarization tensors $e^{+,\times}_{kij}$ satisfying
$e^{+,\times}_{kii}=0$ and $k^{i}e^{+,\times}_{kij}=0$, and the corresponding decomposition is
$\gamma_{kij}(u)=$ $=e^{+}_{kij}\gamma^{+}_{k}(u)+e^{\times}_{kij}\gamma^{\times}_{k}(u)$.
We will denote both mode functions $\gamma^{+,\times}_{k}(u)$ simply as $\gamma_{k}(u)$ in the same way as mode functions of vector perturbations.
The equation for these mode functions with our conventions then can be written as
\begin{eqnarray}
\label{eq:13t2}
\dfrac{d^2\gamma_{k}}{du^2} + \dfrac{2}{u}\dfrac{d\gamma_{k}}{du} + \left( 1 + \dfrac{8}{u^2} \right) \gamma_{k} = 0.
\end{eqnarray}
The analytic solution of this equation is
\begin{eqnarray}
\label{eq:13tt}
\gamma_{k}(u) = u^{-1/2} \left[ c_{11} \textrm{Re}\left\{J_{\sqrt{31}i/2}(u)\right\} + c_{12} \textrm{Re}\left\{Y_{\sqrt{31}i/2}(u)\right\} \right],
\end{eqnarray}
where again the imaginary degree of Bessel functions obstructs our insight into the behavior of such functions.
The approximative behavior of tensor modes is given by
\begin{eqnarray}
\label{eq:13tta}
\gamma_{k}(u) \approx \left\{ \begin{array}{ll}
u^{-1/2} \left[c_{13} \cos\left(\dfrac{1}{2}\sqrt{31}\ln u\right) + c_{14} \sin\left(\dfrac{1}{2}\sqrt{31}\ln u\right) \right], & \textrm{for} \quad u\ll 1, \\
u^{-1} \bigg[c_{15} \cos{u} + c_{16} \sin{u}\bigg], & \textrm{for} \quad u\gg 1.
\end{array} \right.
\end{eqnarray}
Solution (\ref{eq:13tt}) for tensor mode with specific initial conditions is plotted in the right panel of Fig. \ref{fig:03}.
\vskip 2mm
The asymptotic behavior of vector and tensor perturbations is given by
$P^{(n)}_{0}[S]=-3/2$, $P^{(n)}_{0}[\xi_{\perp}]=-1/2$ and $P^{(n)}_{0}[\gamma]=-1/2$
in the superhorizon limit $u\ll 1$, and in the subhorizon limit, $u\gg 1$, we have
$P^{(n)}_{\infty}[S]=-2$, $P^{(n)}_{\infty}[\xi_{\perp}]=0$ and $P^{(n)}_{\infty}[\gamma]=-1$.
Note that in the case with a universe filled with the perfect fluid radiation the solution for perturbation $S_i$, as well as for the vector part of fluid velocity, is proportional to $a^{-2}$,
and the exact solution for modes of tensor perturbations is $\gamma_{k}=u^{-1}\left(\mathcal{C}_1\cos u+\mathcal{C}_2\sin u\right)$.
\vskip 2mm
As we can see, perturbations in a universe filled with the solid-like matter with mater Lagrangian $\mathcal{L}_{\textrm{m}}=-\mathcal{C}\mathcal{X}^2$,
corresponding to $w=1/3$, are not well behaved. There are three problematic issues:
\begin{itemize}
\item Quantities $\widetilde{\mathcal{R}}$ and $\widetilde{\xi}$ describing curvature perturbation
do not coincide with each other and are not conserved in the superhorizon limit.
\item Superluminal sound speed for scalar perturbations in the subhorizon limit, $c_{\textrm{s}}^{(\textrm{S})2}=5/3$.
\item Rapid oscillations of all superhorizon perturbations of the form
$a^{c_{(1)}}\big[c_{(2)}\cos\big(c_{(3)}\ln u\big)+$ $+c_{(4)}\sin\big(c_{(3)}\ln u\big)\big]$ corresponding to infinite sound speed. We will address this in more detail later.
\end{itemize}
In the next section, we relax the radiation-like condition $w=1/3$, and we will study
the case with a more general form of the matter Lagrangian
$\mathcal{L}_{\textrm{m}}=-\mathcal{C}\mathcal{X}^{\mathcal{B}}=-\mathcal{C}\mathcal{X}^{3(w+1)/2}$.
Since this section, dedicated to the special case with $w=1/3$, provided
explanation of conventions being used as well as detailed notes on derivation
of equations for perturbations, the next section with general $w$ will be focused
on differences with respect to the case with $w=1/3$ rather than on writing down
technical details.

\section{Arbitrary pressure to energy density ratio}\label{sec:5}
The background Einstein equations for the universe with flat FLRW metric filled with the matter described by the matter Lagrangian $\mathcal{L}_{\textrm{m}}=-\mathcal{C}\mathcal{X}^{\mathcal{B}}$ are
\begin{eqnarray}
\label{eq:w0_00}
&& \dfrac{\overline{G}_{00}}{8\pi\kappa}=\dfrac{3\mathcal{H}^2}{8\pi\kappa}=\mathcal{C}\left(3\alpha^2\right)^{\mathcal{B}}a^{2(1-\mathcal{B})}=\overline{T}_{00}, \\
\label{eq:w0_ij}
&& \dfrac{\overline{G}_{ij}}{8\pi\kappa}=\dfrac{-\mathcal{H}^2-2\mathcal{H}^{\prime}}{8\pi\kappa}\delta_{ij}=\mathcal{C}\left(3\alpha^2\right)^{\mathcal{B}}\left(-1+\dfrac{2}{3}\mathcal{B}\right)a^{2(1-\mathcal{B})}\delta_{ij}=\overline{T}_{ij}.
\end{eqnarray}
The solution of both equations is
$a(\tau)=\sqrt{3}\left[(\mathcal{B}-1)\sqrt{8\pi\kappa\mathcal{C}}\alpha^{\mathcal{B}}\tau\right]^{1/(\mathcal{B}-1)}\propto\tau^{1/(\mathcal{B}-1)}$.
The scale factor is proportional to the power $1/(\mathcal{B}-1)$, which written through the pressure to energy density ratio
is the standard factor $2/(1+3w)$, the same as for perfect fluid. For $\mathcal{B}>1$ or $w>-1/3$ we will use the convention in which $\tau=0$ corresponds
to the cosmological singularity, $a=0$, and the conformal time is from the interval $\tau\in(0,\infty)$.
For $\mathcal{B}<1$ or $w<-1/3$ we obtain accelerating expansion with a negative power of the conformal time in relation for the scale factor.
In such case we have to use convention in which $\tau\in(-\infty,0)$, and modify the formula for the scale factor,
$a(\tau)=\sqrt{3}\left[(\mathcal{B}-1)\sqrt{8\pi\kappa\mathcal{C}}\alpha^{\mathcal{B}}(-\tau)\right]^{1/(\mathcal{B}-1)}\propto$ $\propto(-\tau)^{1/(\mathcal{B}-1)}$.
The special case with $\mathcal{B}=1$ or $w=-1/3$ will be treated separately in the next section.
\vskip 2mm
Let us now continue the analysis of the solid-like model for linearized perturbations.
With the same parameterization and conventions as before (\ref{eq:mem}),
the perturbed stress-energy tensor is given by
\begin{eqnarray}\label{eq:stressw}
&& \delta T_{00} = \dfrac{3\mathcal{H}^2}{8\pi\kappa} \left[ 2\left(\phi + \mathcal{B}\psi\right) + \dfrac{2}{3} \mathcal{B} \triangle\zeta \right], \quad \delta T_{0i} = \dfrac{3\mathcal{H}^2}{8\pi\kappa} \left[ - S_i + \dfrac{2}{3} \mathcal{B} \left( \zeta_{,i} + \xi_{\perp i} \right)^{\prime} \right], \nonumber\\
&& \delta T_{ij} = \dfrac{3\mathcal{H}^2}{8\pi\kappa} \bigg\{ \left[ \left(2+\dfrac{2}{3}\mathcal{B}(2\mathcal{B}-5)\right)\psi + \dfrac{2}{9}\mathcal{B}(2\mathcal{B}-5)\triangle\zeta \right] \delta_{ij} + \nonumber\\
&& \phantom{\delta T_{ij} = \dfrac{3\mathcal{H}^2}{8\pi\kappa} \bigg\{} + \dfrac{2}{3}\mathcal{B} \left( 2 \zeta_{,ij} + \xi_{\perp i,j} + \xi_{\perp j,i} \right) - \gamma_{ij} \bigg\},
\end{eqnarray}
where we have already used the background solution, so that $\mathcal{H}=\tau^{-1}/(\mathcal{B}-1)$.
This background solution is the only thing that changes the form of the Einstein tensor components.
It is then straightforward to generalize equations (\ref{eq:131})-(\ref{eq:134}), and in the same way as the equation (\ref{eq:13psi})
together with relations (\ref{eq:13zeta}) and (\ref{eq:13phi}) are derived from them, one can derive \begin{eqnarray}
\label{eq:wpsi}
&& \dfrac{d^2\psi_k}{du^2} + 4\dfrac{6\dfrac{2\mathcal{B}-1}{\mathcal{B}-1}+\dfrac{1}{2}\mathcal{B}(\mathcal{B}-1)u^2}{12+(\mathcal{B}-1)^2 u^2}\dfrac{1}{u}\dfrac{d\psi_k}{du} + \nonumber\\
&& \phantom{\dfrac{d^2\psi_k}{du^2}} + \dfrac{1}{3}\dfrac{72\dfrac{\mathcal{B}^2+2\mathcal{B}-1}{(\mathcal{B}-1)^2}+12(5\mathcal{B}-1)u^2+(2\mathcal{B}+1)(\mathcal{B}-1)^2 u^4}{12+(\mathcal{B}-1)^2 u^2}\dfrac{1}{u^2}\psi_k = 0, \\
\label{eq:wzeta}
&& k^2\zeta_k = \dfrac{1}{\mathcal{B}}\dfrac{(\mathcal{B}-1)^2u^2}{12+(\mathcal{B}-1)^2u^2} \left[ (3(\mathcal{B}+1)+(\mathcal{B}-1)^2u^2)\psi_k + 3(\mathcal{B}-1)u \dfrac{d\psi_k}{du} \right], \\
\label{eq:wphi}
&& \phi_k = -\dfrac{3}{12+(\mathcal{B}-1)^2u^2} \left[ (4\mathcal{B}+(\mathcal{B}-1)^2u^2)\psi_k + 4(\mathcal{B}-1)u\dfrac{d\psi_k}{du} \right],
\end{eqnarray}
Similarly, the generalization of relations (\ref{eq:rest1})-(\ref{eq:rest3}) reads
\begin{eqnarray}
\label{eq:rest1w}
&& \delta_k \equiv \dfrac{\widetilde{\delta \rho}_k}{\overline{\rho}} = 2\mathcal{B}\psi_k - \dfrac{2}{3}\mathcal{B}k^2\zeta_k = \dfrac{2}{12+(\mathcal{B}-1)^2u^2} \bigg[ \big(12\mathcal{B}-(\mathcal{B}-1)^2u^2- \nonumber\\
&& \phantom{\delta_k \equiv \dfrac{\widetilde{\delta \rho}_k}{\overline{\rho}} = 2\mathcal{B}\psi_k - \dfrac{2}{3}\mathcal{B}k^2\zeta_k =} -\dfrac{1}{3}(\mathcal{B}-1)^4u^4\big)\psi_k-(\mathcal{B}-1)^3u^3\dfrac{d\psi_k}{du} \bigg], \\
\label{eq:rest2w}
&& \widetilde{\mathcal{R}}_k = -\psi_k - \mathcal{H}\zeta_k^{\prime} = \dfrac{1}{\mathcal{B}}\dfrac{(\mathcal{B}-1)^2u^2}{12+(\mathcal{B}-1)^2u^2} \left[ (3-\mathcal{B})\psi_k - (\mathcal{B}-1)u\dfrac{d\psi_k}{du} \right], \\
\label{eq:rest3w}
&& \widetilde{\xi}_k = -\dfrac{1}{3}k^2\zeta_k = -\dfrac{1}{3\mathcal{B}}\dfrac{(\mathcal{B}-1)^2u^2}{12+(\mathcal{B}-1)^2u^2} \bigg[ \big(3(\mathcal{B}+1)+(\mathcal{B}-1)^2u^2\big)\psi_k + \nonumber\\
&& \phantom{\widetilde{\zeta}_k = -\dfrac{1}{3}k^2\zeta_k =} +3(\mathcal{B}-1)u\dfrac{d\psi_k}{du} \bigg].
\end{eqnarray}
\vskip 2mm
Since equation (\ref{eq:13psi}) is already too complicated to be solved analytically, and (\ref{eq:wpsi}) is its generalization, we have to resort to either numerical solution or superhorizon and subhorizon approximations.
Here we focus on approximative solutions. In the superhorizon limit (\ref{eq:13psisup}) generalizes to
\begin{eqnarray}\label{eq:wpsisup}
\psi_k(u) = \left\{
\begin{array}{ll}
u^{-\lambda} \left[c_{17} u^{|\vartheta|} +c_{18} u^{-|\vartheta|} \right], & \textrm{for} \quad \mathcal{B} < \mathcal{B}_1 \textrm{ or } \mathcal{B} > \mathcal{B}_2, \\
c_{19} u^{-\lambda}, & \textrm{for} \quad \mathcal{B} = \mathcal{B}_1 \textrm{ or } \mathcal{B} = \mathcal{B}_2, \\
u^{-\lambda} \left[c_{20} \cos \left( |\vartheta| \ln u \right) +c_{21} \sin \left( |\vartheta| \ln u \right) \right], & \textrm{for} \quad \mathcal{B} \in (\mathcal{B}_1,\mathcal{B}_2),
\end{array}
\right.
\end{eqnarray}
where $\lambda$ and $\vartheta$ are constants defined as
\begin{eqnarray}
\lambda = \dfrac{3\mathcal{B}-1}{2(\mathcal{B}-1)},
\quad
\vartheta = \dfrac{\sqrt{\mathcal{B}^2-22\mathcal{B}+9}}{2(\mathcal{B}-1)},
\end{eqnarray}
and values $\mathcal{B}_1=11-4\sqrt{7}\dot{=}0.417$ and $\mathcal{B}_2=11+4\sqrt{7}\dot{=}21.6$ determine intervals in the parameter space with two qualitatively different types of the superhorizon solution. However, the second value corresponds to $w_2=(19+8\sqrt{7})/3\dot{=}13.4$, which is far from values for the pressure to energy density ratio within the interval $[-1,1]$, and only the first of the two values of $\mathcal{B}=\mathcal{B}_1$ yields pressure to energy density ratio with the value $w_1=(19-8\sqrt{7})/3\dot{=}-0.722$ which is in this preferred interval.
\vskip 2mm
The consequence of the peculiar form of modes in the third line of (\ref{eq:wpsisup}) for $\mathcal{B}\in(\mathcal{B}_1,\mathcal{B}_2)$ is that the wavefront is given by relation
\begin{eqnarray}\label{eq:wavefront}
|\vartheta|\ln(k|\tau|) - k_i x^i = \textrm{const.},
\end{eqnarray}
and by differentiating this relation we find that the comoving speed at which the wavefront propagates, $|d\vec{x}/d\tau|$, is $|\vartheta|/(k|\tau|)$. This implies not only superluminality but also infinite sound speed in the limit of infinite wavelength to Hubble horizon ratio.
\vskip 2mm
In the subhorizon limit (\ref{eq:13psisub}) generalizes to
\begin{eqnarray}
\psi_k(u) = U^{-\mu} \left[ c_{22} J_{\mu}\left(U\right) + c_{23} Y_{\mu}\left(U\right) \right],
\end{eqnarray}
where
\begin{eqnarray}\label{eq:sound}
\mu = \dfrac{\mathcal{B}+1}{2(\mathcal{B}-1)}, \quad U = \sqrt{\dfrac{2\mathcal{B}+1}{3}} u \equiv c_{\textrm{s}}^{(\textrm{S})} k \tau.
\end{eqnarray}
Here the sound speed squared for subhorizon scalar perturbations is $c_{\textrm{s}}^{(\textrm{S})2}=(2\mathcal{B}+1)/3$, so that superluminality is avoided for $\mathcal{B} \leq 1$ or $w \leq -1/3$.
\vskip 2mm
The asymptotic behavior of scalar perturbations in the superhorizon limit is given by
\begin{eqnarray}\label{eq:asymp0}
P^{(n)}_{0}[\psi] = P^{(n)}_{0}[\delta] & = & \dfrac{1}{2}\left(1-3\mathcal{B} \pm \textrm{Re}\left\{\sqrt{\mathcal{B}^2-22\mathcal{B}+9}\right\}\right) = \nonumber\\ & = & \dfrac{1}{4}\left(-7-9w \pm \sqrt{3}\textrm{Re}\left\{\sqrt{3w^2-38w-29}\right\}\right), \\
\label{eq:asymp0b}
P^{(n)}_{0}[\widetilde{\mathcal{R}}] = P^{(n)}_{0}[\widetilde{\xi}] & = & \dfrac{1}{2}\left(\mathcal{B}-3 \pm \textrm{Re}\left\{\sqrt{\mathcal{B}^2-22\mathcal{B}+9}\right\}\right) = \nonumber\\ & = & \dfrac{1}{4}\left(3w-3 \pm \sqrt{3}\textrm{Re}\left\{\sqrt{3w^2-38w-29}\right\}\right).
\end{eqnarray}
Note that in this limit $\delta\approx 2\mathcal{B}\psi$.
In the subhorizon limit, we have
\begin{eqnarray}\label{eq:asympi}
&& P^{(n)}_{\infty}[\psi]=-\mathcal{B}=-\dfrac{3}{2}(w+1), \quad
P^{(n)}_{\infty}[\delta]=\mathcal{B}-2=\dfrac{1}{2}(3w-1), \nonumber\\
&& P^{(n)}_{\infty}[\widetilde{\mathcal{R}}]=-1, \quad
P^{(n)}_{\infty}[\widetilde{\xi}]=P^{(n)}_{\infty}[\delta].
\end{eqnarray}
\vskip 2mm
The decoupled system of equations for the Fourier modes of vector perturbations generalizing
(\ref{eq:13s3}) and (\ref{eq:13s4}) reads
\begin{eqnarray}
\label{eq:ws3}
&& \dfrac{d^2 S_{k}}{du^2} + \dfrac{2\mathcal{B}}{(\mathcal{B}-1)}\dfrac{1}{u} \dfrac{d S_{k}}{du} + \left(1+\dfrac{2(3\mathcal{B}-1)}{(\mathcal{B}-1)^2u^2}\right)S_k = 0, \\
\label{eq:ws4}
&& k\xi_{\perp k} = -\dfrac{(\mathcal{B}-1)}{2\mathcal{B}} u \left( S_{k} + \dfrac{1}{2} (\mathcal{B}-1) u \dfrac{S_{k}}{du} \right).
\end{eqnarray}
The exact analytic solution of (\ref{eq:ws3}), generalization of (\ref{eq:13ss}), is
\begin{eqnarray}
S_{k}(u) = u^{-\mu} \left[ c_{24} \textrm{Re}\left\{J_{\vartheta}(u)\right\} + c_{25} \textrm{Re}\left\{Y_{\vartheta}(u)\right\} \right],
\end{eqnarray}
and the order of Bessel function is imaginary for $\mathcal{B}\in(\mathcal{B}_1,\mathcal{B}_2)$. Approximative solutions provide a clearer insight into the behavior of this solution.
In the superhorizon limit, it is
\begin{eqnarray}
S_k(u) = \left\{
\begin{array}{ll}
u^{-\mu} \left[c_{26} u^{|\vartheta|} +c_{27} u^{-|\vartheta|} \right], & \textrm{for} \quad \mathcal{B} < \mathcal{B}_1 \textrm{ or } \mathcal{B} > \mathcal{B}_2, \\
c_{28} u^{-\mu}, & \textrm{for} \quad \mathcal{B} = \mathcal{B}_1 \textrm{ or } \mathcal{B} = \mathcal{B}_2, \\
u^{-\mu} \left[c_{29} \cos \left( |\vartheta| \ln u \right) +c_{30} \sin \left( |\vartheta| \ln u \right) \right], & \textrm{for} \quad \mathcal{B} \in (\mathcal{B}_1,\mathcal{B}_2),
\end{array}
\right.
\end{eqnarray}
and in the subhorizon limit, we have
\begin{eqnarray}
S_{k}(u) = u^{-\mu} \left[ c_{31} J_{\mu}(u) + c_{32} Y_{\mu}(u) \right].
\end{eqnarray}
\vskip 2mm
The equation for tensor modes (\ref{eq:13t2}) from the previous section with $\mathcal{B}=1$ is generalized to
\begin{eqnarray}
\label{eq:wt2}
\dfrac{d^2\gamma_{k}}{du^2} + \dfrac{2}{(\mathcal{B}-1)u}\dfrac{d\gamma_{k}}{du} + \left( 1 + \dfrac{4\mathcal{B}}{(\mathcal{B}-1)^2u^2} \right) \gamma_{k} = 0.
\end{eqnarray}
The full analytic solution of this equation is
\begin{eqnarray}
\gamma_{k}(u) = u^{-\nu} \left[ c_{33} \textrm{Re}\left\{J_{\vartheta}(u)\right\} + c_{34} \textrm{Re}\left\{Y_{\vartheta}(u)\right\} \right],
\end{eqnarray}
where
\begin{eqnarray}
\nu = \dfrac{3-\mathcal{B}}{2(\mathcal{B}-1)}.
\end{eqnarray}
In the superhorizon limit, this solution reduces to
\begin{eqnarray}
\gamma_k(u) = \left\{
\begin{array}{ll}
u^{-\nu} \left[c_{35} u^{|\vartheta|} +c_{36} u^{-|\vartheta|} \right], & \textrm{for} \quad \mathcal{B} < \mathcal{B}_1 \textrm{ or } \mathcal{B} > \mathcal{B}_2, \\
c_{37} u^{-\nu}, & \textrm{for} \quad \mathcal{B} = \mathcal{B}_1 \textrm{ or } \mathcal{B} = \mathcal{B}_2, \\
u^{-\nu} \left[c_{38} \cos \left( |\vartheta| \ln u \right) +c_{39} \sin \left( |\vartheta| \ln u \right) \right], & \textrm{for} \quad \mathcal{B} \in (\mathcal{B}_1,\mathcal{B}_2),
\end{array}
\right.
\end{eqnarray}
and in the subhorizon limit, we have
\begin{eqnarray}
\gamma_{k}(u) = u^{-\nu} \left[ c_{40} J_{\nu}(u) + c_{41} Y_{\nu}(u) \right].
\end{eqnarray}
The superhorizon asymptotic behavior of vector and tensor perturbations is given by
\begin{eqnarray}\label{eq:asymp00}
P^{(n)}_{0}[S] & = & \dfrac{1}{2}\left(-\mathcal{B}-1 \pm \textrm{Re}\left\{\sqrt{\mathcal{B}^2-22\mathcal{B}+9}\right\}\right) = \nonumber\\ & = & \dfrac{1}{4}\left(-5-3w \pm \sqrt{3}\textrm{Re}\left\{\sqrt{3w^2-38w-29}\right\}\right), \\
\label{eq:asymp00b}
P^{(n)}_{0}[\xi_{\perp}] = P^{(n)}_{0}[\gamma] & = & \dfrac{1}{2}\left(\mathcal{B}-3 \pm \textrm{Re}\left\{\sqrt{\mathcal{B}^2-22\mathcal{B}+9}\right\}\right) = \nonumber\\ & = & \dfrac{1}{4}\left(3w-3 \pm \sqrt{3}\textrm{Re}\left\{\sqrt{3w^2-38w-29}\right\}\right),
\end{eqnarray}
with the superluminal propagation speed for all vector and tensor perturbations, $c_{\textrm{s}}=$ \\ $=|\vartheta|/(k|\tau|)$, the same as for scalar perturbations. In the subhorizon limit, we have
\begin{eqnarray}\label{eq:asympii}
P^{(n)}_{\infty}[S] = -\mathcal{B}=-\dfrac{3}{2}(w+1), \quad P^{(n)}_{\infty}[\xi_{\perp}]=\left\{\begin{array}{cl} -1 & \mathcal{B} \neq2 \\ 0 & \mathcal{B}=2 \end{array}\right., \quad P^{(n)}_{\infty}[\gamma]=-1,
\end{eqnarray}
with the sound speed equal to the speed of light for all vector and tensor perturbations.

\section{Singular cases}\label{sec:6}
There are two special cases that have to be studied separately.
We start with the case with $\mathcal{B}=1$ or $w=-1/3$.
The background Einstein equations are
\begin{eqnarray}
\label{eq:b0_00}
&& \dfrac{\overline{G}_{00}}{8\pi\kappa}=\dfrac{3\mathcal{H}^2}{8\pi\kappa}=3\mathcal{C}\alpha^2=\overline{T}_{00}, \\
\label{eq:b0_ij}
&& \dfrac{\overline{G}_{ij}}{8\pi\kappa}=\dfrac{-\mathcal{H}^2-2\mathcal{H}^{\prime}}{8\pi\kappa}\delta_{ij}=-\mathcal{C}\alpha^2\delta_{ij}=\overline{T}_{ij},
\end{eqnarray}
with the solution $a=a_{*}e^{\beta(\tau-\tau_{*})}$, where $\beta=\sqrt{8\pi\kappa\mathcal{C}}\alpha$.
The decoupled system of equations for scalar perturbations is
\begin{eqnarray}
\label{eq:bpsi}
&& \dfrac{d^2\psi_k}{du^2} + 2\sigma\dfrac{d\psi_k}{du} + \left(4\sigma^2+1\right)\psi_k = 0, \\
\label{eq:bzeta}
&& k^2\zeta_k = \dfrac{1}{12\sigma^2+1} \left[ \left(6\sigma^2+1\right)\psi_k + 3\sigma \dfrac{d\psi_k}{du} \right], \\
\label{eq:bphi}
&& \phi_k = \dfrac{1}{12\sigma^2+1} \left[ -3\left(4\sigma^2+1\right)\psi_k - 12\sigma \dfrac{d\psi_k}{du} \right],
\end{eqnarray}
where $\sigma=\beta/k$,
and other important scalar perturbations can be written as
\begin{eqnarray}
\label{eq:rest1b}
&& \delta_k \equiv \dfrac{\widetilde{\delta \rho}_k}{\overline{\rho}} = 2\psi_k - \dfrac{2}{3}k^2\zeta_k = \dfrac{1}{12\sigma^2+1} \bigg[ \dfrac{4}{3}\left(15\sigma^2+1\right)\psi_k-2\sigma\dfrac{d\psi_k}{du} \bigg], \\
\label{eq:rest2b}
&& \widetilde{\mathcal{R}}_k = -\psi_k - \beta\zeta_k^{\prime} = \dfrac{1}{12\sigma^2+1} \left[ 2\psi_k - \dfrac{1}{\sigma}\dfrac{d\psi_k}{du} \right], \\
\label{eq:rest3b}
&& \widetilde{\zeta}_k = -\dfrac{1}{3}k^2\zeta_k = -\dfrac{1}{3}\dfrac{1}{12\sigma^2+1} \bigg[ \big(6\sigma^2+1)\psi_k + 3\sigma\dfrac{d\psi_k}{du} \bigg].
\end{eqnarray}
Decoupled equations for vector perturbations are
\begin{eqnarray}
\label{eq:bs3}
&& \dfrac{d^2 S_{k}}{du^2} + 2\sigma \dfrac{d S_{k}}{du} + \left(4\sigma^2+1\right)S_k = 0, \\
\label{eq:bs4}
&& k\xi_{\perp k} = -\dfrac{1}{2\sigma^2} \left( \sigma S_{k} + \dfrac{1}{2} \dfrac{S_{k}}{du} \right),
\end{eqnarray}
and for tensor perturbations we have
\begin{eqnarray}
\label{eq:bt2}
\dfrac{d^2\gamma_{k}}{du^2} + 2\sigma\dfrac{d\gamma_{k}}{du} + \left( 4\sigma^2+1 \right) \gamma_{k} = 0.
\end{eqnarray}
Modes for scalar perturbation $\psi$, vector perturbations $S_i$, and tensor perturbations $\gamma_{ij}$ obey the same equation with the general solution of the form
\begin{eqnarray}
\chi_k(\tau)=e^{-\beta\tau}\left[c_{42}\cos\left(\sqrt{3\beta^2+k^2}\tau\right)+c_{43}\sin\left(\sqrt{3\beta^2+k^2}\tau\right)\right].
\end{eqnarray}
Therefore, all perturbations decay as $a^{-1}$, which is in agreement with the results of the previous section with the limit $\mathcal{B}\to 1$ taken.
\vskip 2mm
The second singular case is for $\mathcal{B}=0$ or $w=-1$. However, in such case equations for scalar and vector perturbations imply that they have to vanish, and modes of tensor perturbations obey the equation
\begin{eqnarray}
\dfrac{d^2\gamma_{k}}{du^2} - \dfrac{2}{u}\dfrac{d\gamma_{k}}{du} + k^2\gamma_{k} = 0.
\end{eqnarray}
As expected, this case does not differ from the case with a universe with dark energy with $w=-1$ being its only matter component, or alternatively, an empty universe with the cosmological constant, because the matter Lagrangian $\mathcal{L}_{\textrm{m}}=-\mathcal{C}\mathcal{X}^{\mathcal{B}}$ is constant for $\mathcal{B}=0$.
This is obvious also from relations for perturbed energy density and pressure (\ref{eq:pom1}) and the discussion in the following paragraphs in section \ref{sec:3}.

\section{Results and conclusion}\label{sec:7}
\begin{figure}[!htb]
\sbox0{
\includegraphics[scale=0.35]{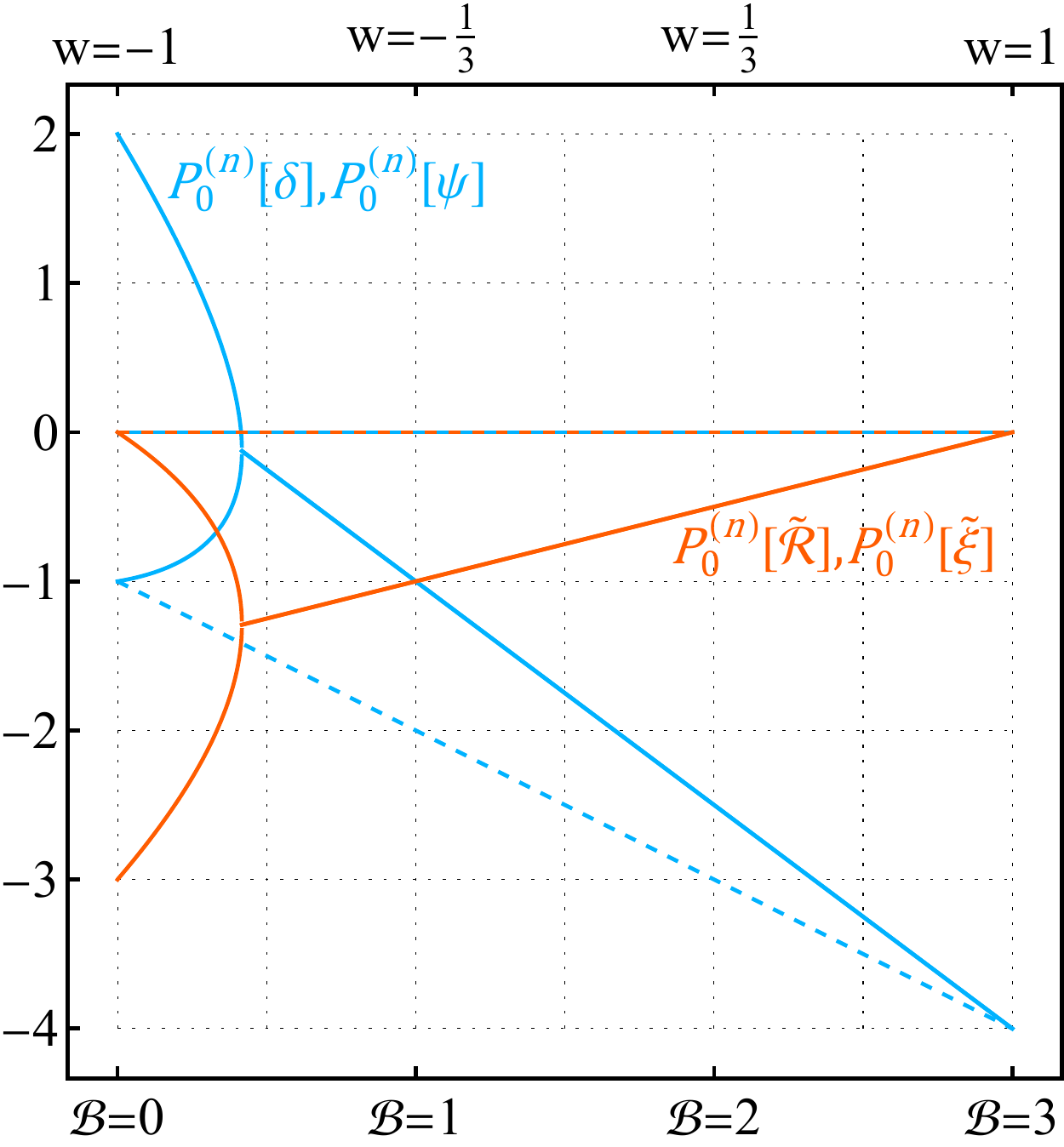}
\includegraphics[scale=0.35]{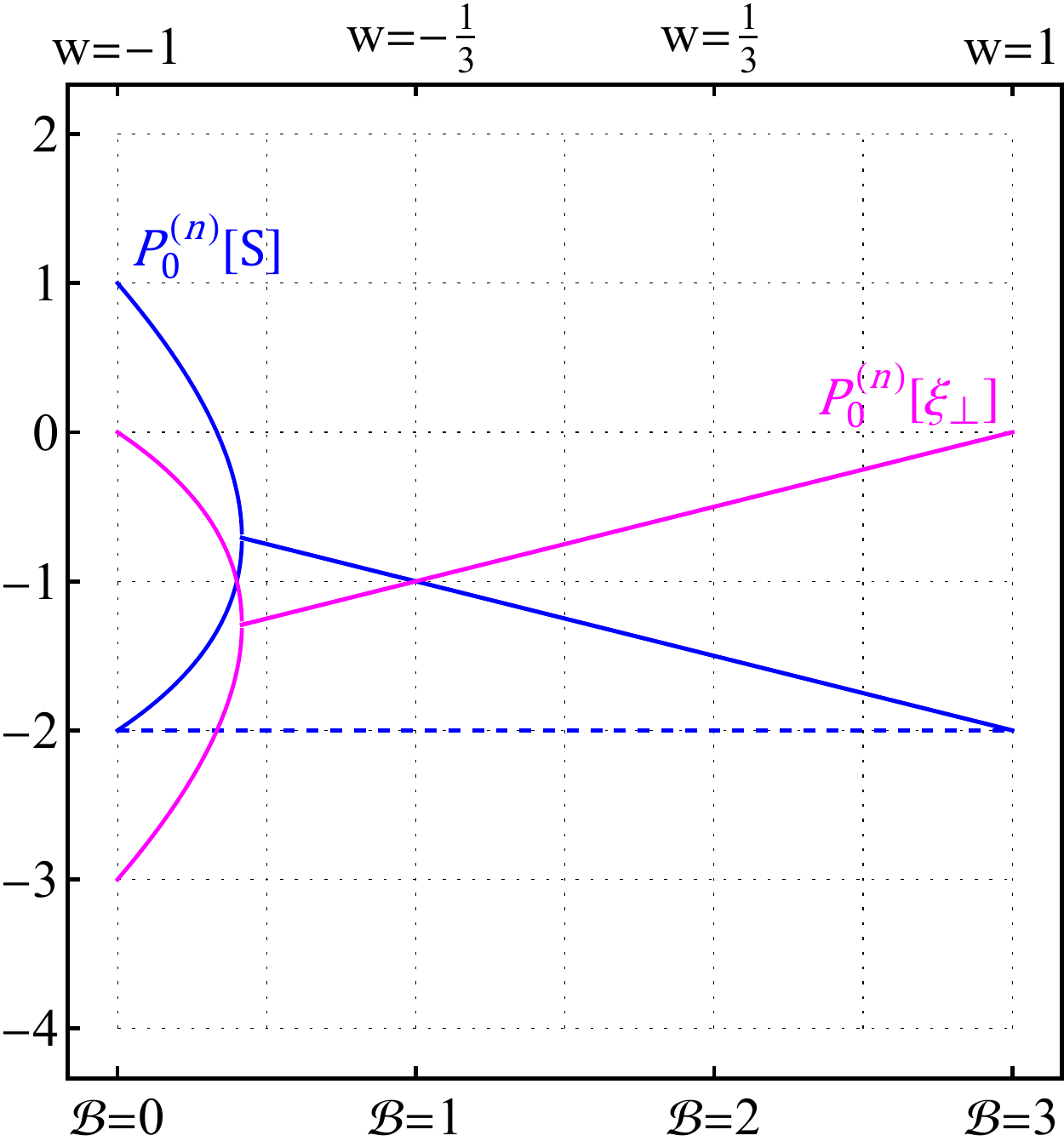}
\includegraphics[scale=0.35]{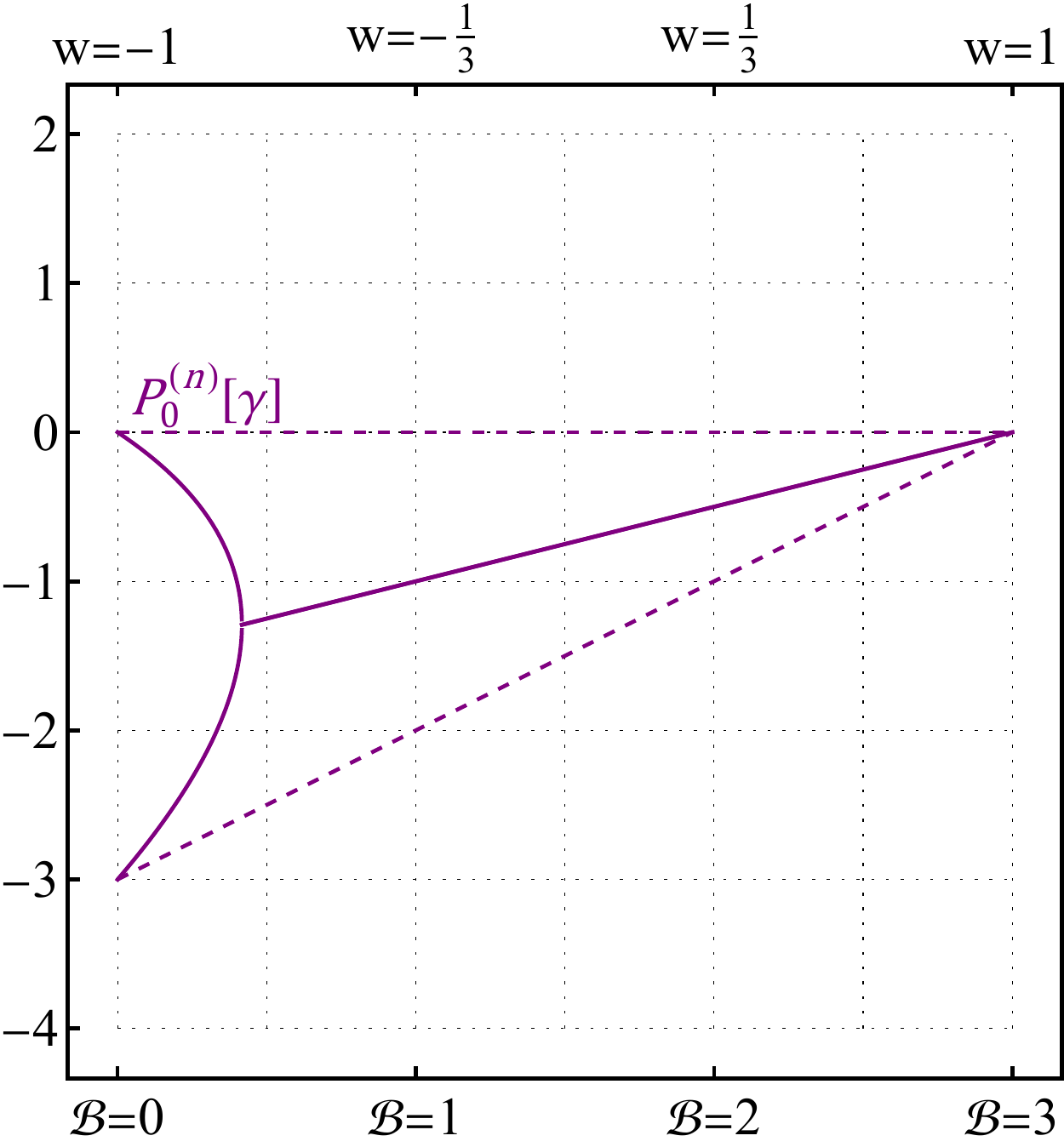}
}
\centering
\begin{minipage}{\wd0}
\usebox0
\linespread{1}
\setlength{\abovecaptionskip}{-2pt plus 0pt minus 0pt}
\caption{\label{fig:04}{\footnotesize
Dependence of quantities (\ref{eq:asymp0}), (\ref{eq:asymp0b}), (\ref{eq:asymp00}) and (\ref{eq:asymp00b}) defined by (\ref{eq:defi1}) and (\ref{eq:defi2}) on parameter $\mathcal{B}=3(w+1)/2$. Solid lines represent the model studied in section \ref{sec:5} and dashed lines represent the case with perfect fluid, where $P^{(1)}_{0}[\psi]=P^{(1)}_{0}[\delta]=0$, $P^{(2)}_{0}[\psi]=P^{(2)}_{0}[\delta]=-(5+3w)/2$, $P^{(n)}_{0}[\widetilde{\mathcal{R}}]=P^{(n)}_{0}[\widetilde{\xi}]=0$, $P^{(n)}_{0}[S]=-2$, $P^{(1)}_{0}[\gamma]=0$, $P^{(2)}_{0}[\gamma]=3(w-1)/2$. {\bf Left panel}: Scalar perturbations, $\phi$ and $\delta$ (light blue) and curvature perturbations $\widetilde{\mathcal{R}}$ and $\widetilde{\xi}$ (orange line). {\bf Middle panel}: Vector perturbations $S_i$ (magenta) and $\xi_{\perp i}$ (blue). {\bf Right panel}: Tensor perturbations $\gamma_{ij}$ (purple). 
}}
\end{minipage}
\end{figure}

We have studied one parametric set of models with the triplet of matter fields $\varphi^i$ in the flat FLRW universe. Assuming that the matter Lagrangian $\mathcal{L}_{\textrm{m}}$ depends only on quantity $\mathcal{X}=g^{\mu\nu}\varphi^{i}_{\phantom{i}\mu}\varphi^{i}_{\phantom{i}\nu}$, the condition of constant pressure to energy density ratio $w$ allows it to be of the form $\mathcal{L}_{\textrm{m}}=-\mathcal{C}\mathcal{X}^{\mathcal{B}}$, where $\mathcal{B}=3(w+1)/2$.
\vskip 2mm
Size of the $(n)$-th mode of arbitrary superhorizon perturbation $\chi$ depends on the scale factor as $a^{P^{(n)}_{0}[\chi]}$, where the power $P^{(n)}_{0}[\chi]$ defined by relations (\ref{eq:defi1}) and (\ref{eq:defi2}) is a constant which is given by the parameter $\mathcal{B}$. The dependence of these power factors on parameter $\mathcal{B}$ or $w=-1+2\mathcal{B}/3$ for all kinds of perturbations is plotted in Fig. \ref{fig:04}.
\vskip 2mm
There are two qualitatively different regimes of the evolution of superhorizon perturbations within the interval for the pressure to energy density ratio $-1< w\leq 1$. If $w>w_1=(19-8\sqrt{7})/3\dot{=}-0.722$, the superhorizon modes are of the form
\begin{eqnarray}
\chi_k(\tau)=|\tau|^{c_{(a)}}\left\{c_{(b)}\cos\left[c_{(c)}\ln\left(k|\tau|\right)\right]+c_{(d)}\sin\left[c_{(c)}\ln\left(k|\tau|\right)\right]\right\}.
\end{eqnarray}
This implies superluminal sound speed $c_{\textrm{s}}=c_{(c)}/(k|\tau|)$ for superhorizon modes with small enough $|u|=|k\tau|$, which also diverges in the limit of infinite wavelength to Hubble horizon ratio. We will address this issue in more detail later in this section. For $w\leq w_1$ the superhorizon modes are better behaved with two independent modes of the form of power functions of the scale factor, $\chi_{k}(\tau)\propto a^{\textrm{const.}}$. The same is true also for $w\geq w_2\equiv (19+$ $+8\sqrt{7})/3\dot{=}13.4$, but the value of $w_2$ is far from the interval $[-1,1]$, which is the usual range of values considered for pressure to energy density ratio. One can see the change of behavior at $w=w_1$ in all three panels of Fig. \ref{fig:04}, for scalar, vector, and tensor perturbations.
\vskip 2mm
This behavior differs from the case with perfect fluid, where all scalar and tensor perturbations have one constant mode and one decaying mode in the superhorizon limit, while all vector perturbations decay. Another distinctive feature of more standard models is the conservation of superhorizon modes of scalar quantities $\widetilde{\mathcal{R}}$ and $\widetilde{\xi}$ parameterizing curvature perturbation. In our model, their size depends on the scale factor as a power function with the power given by (\ref{eq:asymp0b}). This is depicted by the solid orange line in the first panel of Fig. \ref{fig:04}. Moreover, unlike in the case with perfect fluid or other simpler models, in our model quantities $\widetilde{\mathcal{R}}$ and $\widetilde{\xi}$ are not equal in the superhorizon limit, but at least their asymptotic behavior is the same.
\vskip 2mm
The important issue that needs to be explained in more detail concerns the superluminality of the sound speed. For scalar perturbations in the subhorizon limit, it is given by $c_{\textrm{s}}^{(\textrm{S})2}=(2\mathcal{B}+1)/3=w+4/3$, see (\ref{eq:sound}), while for vector and tensor perturbations the speed of their propagation equals the speed of light, $c_{\textrm{s}}^{(\textrm{V})}=c_{\textrm{s}}^{(\textrm{T})}=1$. This result is in agreement with the model of solid inflation \cite{gruzinov,endlich,akhshik,bartolo,bordin} with the more general form of the matter Lagrangian $\mathcal{L}_{\textrm{m}}=F(\mathcal{X},\mathcal{Y},\mathcal{Z})$, where $\mathcal{X}$ is defined in the same way as in our paper in (\ref{eq:lag}), and additional quantities are defined as $\mathcal{Y}=\textrm{Tr}(B^2)/\mathcal{X}^2$ and $\mathcal{Z}=\textrm{Tr}(B^3)/\mathcal{X}^3$ with components of the body metric defined as $B^{ij}=g^{\mu\nu}\varphi^{i}_{\phantom{i},\mu}\varphi^{j}_{\phantom{j},\nu}$, so that quantity $\mathcal{X}$ is simply trace of the body metric $\mathcal{X}=\textrm{Tr}(B)$. In this model, perturbations propagate with speeds of sound given by relations
\begin{eqnarray}
c_{\textrm{s}}^{(\textrm{S})2}=1+\dfrac{2}{3}\dfrac{\mathcal{X}\partial^2_{
\mathcal{X}}F}{\partial_{\mathcal{X}}F}+\dfrac{8}{9}\dfrac{\partial_{\mathcal{Y}}F+\partial_{\mathcal{Z}}F}{\mathcal{X}\partial_{\mathcal{X}}F}, \quad c_{\textrm{s}}^{(\textrm{V})2}=c_{\textrm{s}}^{(\textrm{T})2}=1+\dfrac{2}{3}\dfrac{\partial_{\mathcal{Y}}F+\partial_{\mathcal{Z}}F}{\mathcal{X}\partial_{\mathcal{X}}F}.
\end{eqnarray}
The model studied in this paper corresponds to $F\propto\mathcal{X}^{\mathcal{B}}$, and therefore, we indeed obtain $c_{\textrm{s}}^{(\textrm{S})2}=1+(2/3)(\mathcal{B}-1)=(2\mathcal{B}+1)/3$ and $c_{\textrm{s}}^{(\textrm{V})2}=c_{\textrm{s}}^{(\textrm{T})2}=1$.
Note that in the model with matter Lagrangian given by $\mathcal{L}_{\textrm{m}}=F(\mathcal{X},\mathcal{Y},\mathcal{Z})$, the pressure to energy density ratio $\overline{w}=\overline{p}/\overline{\rho}$ remains of the form (\ref{eq:wratio}).
Superluminality then can be avoided much easier by adjusting the factor $(\partial_{\mathcal{Y}}F+\partial_{\mathcal{Z}}F)/\mathcal{X}\partial_{\mathcal{X}}F$, however, such freedom is restricted by the constraint $3c_{\textrm{s}}^{(\textrm{S})2}-4c_{\textrm{s}}^{(\textrm{V/T})2}=3\overline{w}-(\overline{w}+1)^{-1}(d\overline{w}/d\textrm{ln}a)$.
\vskip 2mm
In order to avoid superluminality and instability of scalar perturbations in the subhorizon limit, one has to demand the condition $0\leq c_{\textrm{s}}^{(\textrm{S})2} \leq 1$. It is satisfied for $-1/2\leq\mathcal{B}\leq 1$ or $-4/3\leq w\leq -1/3$. Hence, either the parameter $\mathcal{B}$ is allowed to be from only the mentioned interval or for values of $\mathcal{B}$ outside of this interval there is some mechanism that prevents the formation of subhorizon perturbations. One such physical mechanism may be cosmic inflation occurring before the era during which the universe can be described by the model studied in this paper. During inflation even quantum fluctuations with Planckian wavelength size may be stretched to superhorizon scale, and we can take into account cases with $\mathcal{B}>1$ or $w>-1/3$ as well, in spite of superluminal sound speed in the subhorizon limit.
\vskip 2mm
However, due to superluminality concerning also perturbations in the superhorizon limit, the parameter space has to be restricted regardless of inflationary stretching. From the relation for the wavefront of superhorizon modes (\ref{eq:wavefront}) valid for $w\in(w_1,w_2)$ we have derived the speed of the wavefront propagation $c_{\textrm{s}} = |d\vec{x}/d\tau| = |\vartheta|/(k|\tau|)$, which can be rewritten as
\begin{eqnarray}
c_{\textrm{s}} = \dfrac{1}{k|\tau|}\dfrac{\textrm{Im}\sqrt{\mathcal{B}^2-22\mathcal{B}+9}}{2|\mathcal{B}-1|} = \dfrac{\sqrt{3}}{2} \dfrac{1}{k|\tau|} \dfrac{\textrm{Im}\sqrt{3w^2-38w-29}}{|3w+1|}.
\end{eqnarray}
This problem with superhorizon superluminality does not occur if $\mathcal{B}\leq \mathcal{B}_1=11-4\sqrt{7}\dot{=}0.417$, $w\leq w_1=(19-8\sqrt{7})/3\dot{=}-0.722$ or $\mathcal{B}\geq\mathcal{B}_2=11+4\sqrt{7}\dot{=}21.6$, $w\geq w_2(19+8\sqrt{7})/3\dot{=}13.4$
\vskip 2mm
In conclusion, the allowed region of the parameter space of the model studied in this paper is given by the interval $\mathcal{B}\in[0,\mathcal{B}_1]$ corresponding to the interval for pressure to energy density ratio $w\in[-1,w_1]$. This interval can in principle be extended to $\mathcal{B}\in[-1/2,\mathcal{B}_1]$ or $w\in[-4/3,w_1]$, but the pressure to energy density ratio smaller than $-1$ leads to big rip, a divergence of the scale factor at some finite time. Hence, the era described by our model with $\mathcal{B}\in[-1/2,0)$ or $w\in[-4/3,0)$ should not last long enough to reach the big rip. Note also that in the case with $\mathcal{B}=0$ or $w=-1$ no scalar and vector perturbations can be formed, and tensor perturbations evolve in the same way as in the case with perfect fluid or with the cosmological constant in absence of other kinds of matter.
\vskip 2mm
Even with the restriction on the region of the parameter space mentioned above taken into account, the behavior of superhorizon perturbations is qualitatively different from models with perfect fluid. While there is one constant mode and one decaying mode for scalar and tensor superhorizon perturbations and vector perturbations decay in perfect fluid models, both independent superhorizon modes are power functions of the scale factor in our model for allowed values of parameter $\mathcal{B}$. As we can see in Fig. \ref{fig:04}, there are superhorizon modes of scalar perturbations $\delta$ and $\psi$ and vector perturbations $S_i$ which can grow in the course of the expansion of the universe. There is a growing mode of $\delta$ as well as $\psi$ for $\mathcal{B}<\sqrt{2}-1\dot{=}0.414$ or $w<(2\sqrt{2}-5)/3=-0.724$, and one of the modes of perturbation $S_i$ grows for $\mathcal{B}<1/3$ or $w<-7/9$.
\vskip 2mm
In the case with perfect fluid all superhorizon scalar and tensor perturbations are dominated by the nondecaying constant part, and therefore, the tensor to scalar ratio is conserved. In our model, this is true only if we define it through curvature perturbations,
\begin{eqnarray}
r_{(1)}=\mathcal{O}(\tau)\lim\limits_{k|\tau|\to 0}\dfrac{(\gamma_k(\tau))^2}{(\widetilde{\mathcal{R}}_k(\tau))^2}, \quad r_{(2)}=\mathcal{O}(\tau)\lim\limits_{k|\tau|\to 0}\dfrac{(\gamma_k(\tau))^2}{(\widetilde{\xi}_k(\tau))^2},
\end{eqnarray}
where $\mathcal{O}(\tau)$ denotes functions that either oscillate with constant amplitude or are constant, defined so that both $r_{(1)}$ and $r_{(2)}$ are constant. The reason is that $P^{(n)}_0[\gamma]=P^{(n)}_0[\widetilde{\mathcal{R}}]=$ $=P^{(n)}_0[\widetilde{\xi}]$.
If we define this quantity through gauge invariant metric perturbation $\psi$ or invariant fractional energy density perturbation $\delta$,
\begin{eqnarray}
r_{(3)}(\tau)=\mathcal{O}(\tau)\lim\limits_{k|\tau|\to 0}\dfrac{(\gamma_k(\tau))^2}{(\psi_k(\tau))^2}=2\mathcal{B}\mathcal{O}(\tau)\lim\limits_{k|\tau|\to 0}\dfrac{(\gamma_k(\tau))^2}{(\delta_k(\tau))^2},
\end{eqnarray}
we obtain a function with power law dependence on the scale factor, $r_{(3)}(\tau)\propto a^{4(\mathcal{B}-1)}=$ $=a^{2(3w+1)}$. This means that tensor to scalar ratio of superhorizon perturbations defined in the second way is constant, like in ordinary models, only for $\mathcal{B}=1$ or $w=-1/3$.
Similarly, one can define also vector to scalar ratio in two ways
\begin{eqnarray}
s_{(1)}(\tau)=\mathcal{O}(\tau)\lim\limits_{k|\tau|\to 0}\dfrac{(S_k(\tau))^2}{(\widetilde{\mathcal{R}}_k(\tau))^2}, \quad s_{(2)}=\mathcal{O}(\tau)\lim\limits_{k|\tau|\to 0}\dfrac{(S_k(\tau))^2}{(\widetilde{\xi}_k(\tau))^2},
\end{eqnarray}
and
\begin{eqnarray}
s_{(3)}(\tau)=\mathcal{O}(\tau)\lim\limits_{k|\tau|\to 0}\dfrac{(S_k(\tau))^2}{(\psi_k(\tau))^2}=2\mathcal{B}\mathcal{O}(\tau)\lim\limits_{k|\tau|\to 0}\dfrac{(S_k(\tau))^2}{(\delta_k(\tau))^2}.
\end{eqnarray}
The dependence of these quantities on the scale factor is $s_{(1)}(\tau)=s_{(2)}(\tau)\propto a^{-2(\mathcal{B}-1)}=$ $=a^{-(3w+1)}$ and $s_{(3)}(\tau)\propto a^{2(\mathcal{B}-1)}=a^{3w+1}$. Note that in the case with perfect fluid or any matter with zero shear components of the stress-energy tensor all quantities defined above are proportional to $a^{-4}$, since $S_i\propto a^{-2}$.
\vskip 2mm
In summary, in the restricted region of the parameter space, for $\mathcal{B}<\mathcal{B}_1$, we have
\begin{eqnarray}
\dfrac{d\ln s_{(1)}}{d\ln a}=\dfrac{d\ln s_{(2)}}{d\ln a}>\dfrac{d\ln r_{(1)}}{d\ln a}=\dfrac{d\ln r_{(2)}}{d\ln a}>\dfrac{d\ln s_{(3)}}{d\ln a}>\dfrac{d\ln r_{(3)}}{d\ln a},
\end{eqnarray}
since $-2(\mathcal{B}-1)>0>2(\mathcal{B}-1)>4(\mathcal{B}-1)$ for any $\mathcal{B}<1$. If we follow more standard definitions of the tensor to scalar ratio and vector to scalar ratio through curvature perturbations, i.e. we disregard quantities $r_{(3)}$ and $s_{(3)}$, we may conclude that tensor to scalar ration remains constant, while vector to scalar ratio grows. This result considerably differs from predictions of simpler models like single field models lacking vector perturbations or perfect fluid models which predict decreasing vector to scalar ratio.

\section*{Acknowledgements}
The work was supported by grants VEGA 1/0719/23, VEGA 1/0025/23, and Ministry of Education contract No. 0466/2022.

\appendix
\renewcommand{\thesection}{\Alph{section}}

{\setstretch{1.0}

}

\end{document}